\documentclass[12pt,a4paper]{article}
\pdfoutput=1

\usepackage{ifthen} 
\newboolean{pdflatex}
\setboolean{pdflatex}{true} 

\newboolean{articletitles}
\setboolean{articletitles}{true} 

\newboolean{uprightparticles}
\setboolean{uprightparticles}{false} 

\def\paperauthors{LHCb collaboration} 
\def\papertitle{A search for \\ \XiccppDecayDp decays} 
\def\papercopyright{\the\year\ CERN for the benefit of the LHCb collaboration} 
\def\paperlicence{CC-BY-4.0 licence}
\def\paperlicenceurl{https://creativecommons.org/licenses/by/4.0/}

\usepackage[top=1in, bottom=1.25in, left=1in, right=1in]{geometry}

\usepackage{microtype}
\usepackage{lineno}  
\usepackage{xspace} 
\usepackage{caption} 

\usepackage{graphicx}  
\usepackage{color}
\usepackage{colortbl}
\graphicspath{{./figs/}} 

\usepackage{tabularx}
\usepackage{dcolumn}
\newcolumntype{d}[1]{D{.}{.}{#1}}

\usepackage{amsmath} 
\usepackage{amssymb}
\usepackage{amsfonts}
\usepackage{upgreek} 

\newcommand*\patchAmsMathEnvironmentForLineno[1]{%
\expandafter\let\csname old#1\expandafter\endcsname\csname #1\endcsname
\expandafter\let\csname oldend#1\expandafter\endcsname\csname
end#1\endcsname
 \renewenvironment{#1}%
   {\linenomath\csname old#1\endcsname}%
   {\csname oldend#1\endcsname\endlinenomath}%
}
\newcommand*\patchBothAmsMathEnvironmentsForLineno[1]{%
  \patchAmsMathEnvironmentForLineno{#1}%
  \patchAmsMathEnvironmentForLineno{#1*}%
}
\AtBeginDocument{%
\patchBothAmsMathEnvironmentsForLineno{equation}%
\patchBothAmsMathEnvironmentsForLineno{align}%
\patchBothAmsMathEnvironmentsForLineno{flalign}%
\patchBothAmsMathEnvironmentsForLineno{alignat}%
\patchBothAmsMathEnvironmentsForLineno{gather}%
\patchBothAmsMathEnvironmentsForLineno{multline}%
}

\usepackage{hyperref}    
\usepackage[all]{hypcap} 

\usepackage{xspace} 
\usepackage{upgreek}

\def\lhcb   {\mbox{LHCb}\xspace}

\def\MagUp {\mbox{\em Mag\kern -0.05em Up}\xspace}

\ifthenelse{\boolean{uprightparticles}}%
{

 \def\Ppi         {\ensuremath{\uppi}\xspace}

 \def\PDelta      {\ensuremath{\Delta}\xspace}                 
 \def\PXi         {\ensuremath{\Xi}\xspace}                 
 \def\PLambda     {\ensuremath{\Lambda}\xspace}                 
 \def\PSigma      {\ensuremath{\Sigma}\xspace}                 
 \def\POmega      {\ensuremath{\Omega}\xspace}                 
 \def\PUpsilon    {\ensuremath{\Upsilon}\xspace}

 \def\PB      {\ensuremath{\mathrm{B}}\xspace}                 
                  
 \def\PD      {\ensuremath{\mathrm{D}}\xspace}

 \def\PK      {\ensuremath{\mathrm{K}}\xspace}

 \def\Pb      {\ensuremath{\mathrm{b}}\xspace}                 
 \def\Pc      {\ensuremath{\mathrm{c}}\xspace}

 \def\Pi      {\ensuremath{\mathrm{i}}\xspace}

 \def\Pp      {\ensuremath{\mathrm{p}}\xspace}

 \def\thebaroffset{0.0em}
}
{

 \def\Ppi         {\ensuremath{\pi}\xspace}

 \mathchardef\PDelta="7101
 \mathchardef\PXi="7104
 \mathchardef\PLambda="7103
 \mathchardef\PSigma="7106
 \mathchardef\POmega="710A
 \mathchardef\PUpsilon="7107
                  
 \def\PB      {\ensuremath{B}\xspace}                 
                  
 \def\PD      {\ensuremath{D}\xspace}

 \def\PK      {\ensuremath{K}\xspace}

 \def\Pb      {\ensuremath{b}\xspace}                 
 \def\Pc      {\ensuremath{c}\xspace}

 \def\Pi      {\ensuremath{i}\xspace}

 \def\Pp      {\ensuremath{p}\xspace}

 \def\thebaroffset{0.18em}
}
\newcommand{\offsetoverline}[2][\thebaroffset]{\kern #1\overline{\kern -#1 #2}}%

\makeatletter
\ifcase \@ptsize \relax
  \newcommand{\miniscule}{\@setfontsize\miniscule{4}{5}}
\or
  \newcommand{\miniscule}{\@setfontsize\miniscule{5}{6}}
\or
  \newcommand{\miniscule}{\@setfontsize\miniscule{5}{6}}
\fi
\makeatother

\DeclareRobustCommand{\optbar}[1]{\shortstack{{\miniscule (\rule[.5ex]{1.25em}{.18mm})}
  \\ [-.7ex] $#1$}}

\def\cquark    {{\ensuremath{\Pc}}\xspace}

\def\bquark    {{\ensuremath{\Pb}}\xspace}

\def\pion   {{\ensuremath{\Ppi}}\xspace}

\def\pip    {{\ensuremath{\pion^+}}\xspace}
\def\pim    {{\ensuremath{\pion^-}}\xspace}

\def\kaon    {{\ensuremath{\PK}}\xspace}

\def\KorKbar {\kern \thebaroffset\optbar{\kern -\thebaroffset \PK}{}\xspace}

\def\Kp      {{\ensuremath{\kaon^+}}\xspace}
\def\Km      {{\ensuremath{\kaon^-}}\xspace}

\def\D       {{\ensuremath{\PD}}\xspace}

\def\DorDbar {\kern \thebaroffset\optbar{\kern -\thebaroffset \PD}\xspace}

\def\Dp      {{\ensuremath{\D^+}}\xspace}

\def\BorBbar {\kern \thebaroffset\optbar{\kern -\thebaroffset \PB}\xspace}

\def\Y#1S{\ensuremath{\PUpsilon{(#1S)}}\xspace}

\def\proton      {{\ensuremath{\Pp}}\xspace}

\def\LorLbar     {\kern \thebaroffset\optbar{\kern -\thebaroffset \PLambda}\xspace}
\def\Lambdares   {{\ensuremath{\PLambda}}\xspace}

\def\Sigmares    {{\ensuremath{\PSigma}}\xspace}

\def\Xires       {{\ensuremath{\PXi}}\xspace}

\def\Xic         {{\ensuremath{\Xires_\cquark}}\xspace}

\def\Xicc        {{\ensuremath{\Xires_{\cquark\cquark}}}\xspace}

\def\Xiccp       {{\ensuremath{\Xires^+_{\cquark\cquark}}}\xspace}
\def\Xiccpp      {{\ensuremath{\Xires^{++}_{\cquark\cquark}}}\xspace}

\def\to                 {\ensuremath{\rightarrow}\xspace}

\def\AT#1     {\ensuremath{A_{\mathrm{T}}^{#1}}\xspace}           

\def\C#1      {\ensuremath{\mathcal{C}_{#1}}\xspace}                       
\def\Cp#1     {\ensuremath{\mathcal{C}_{#1}^{'}}\xspace}                    
\def\Ceff#1   {\ensuremath{\mathcal{C}_{#1}^{\mathrm{(eff)}}}\xspace}        
\def\Cpeff#1  {\ensuremath{\mathcal{C}_{#1}^{'\mathrm{(eff)}}}\xspace}       
\def\Ope#1    {\ensuremath{\mathcal{O}_{#1}}\xspace}                       
\def\Opep#1   {\ensuremath{\mathcal{O}_{#1}^{'}}\xspace}                    

\newcommand{\nospaceunit}[1]{\ensuremath{\text{#1}}}       
\newcommand{\aunit}[1]{\ensuremath{\text{\,#1}}}       

\newcommand{\tev}{\aunit{Te\kern -0.1em V}\xspace}
\newcommand{\gev}{\aunit{Ge\kern -0.1em V}\xspace}
\newcommand{\mev}{\aunit{Me\kern -0.1em V}\xspace}
\newcommand{\kev}{\aunit{ke\kern -0.1em V}\xspace}
\newcommand{\ev}{\aunit{e\kern -0.1em V}\xspace}
\newcommand{\mevc}{\ensuremath{\aunit{Me\kern -0.1em V\!/}c}\xspace}
\newcommand{\gevc}{\ensuremath{\aunit{Ge\kern -0.1em V\!/}c}\xspace}
\newcommand{\mevcc}{\ensuremath{\aunit{Me\kern -0.1em V\!/}c^2}\xspace}
\newcommand{\gevcc}{\ensuremath{\aunit{Ge\kern -0.1em V\!/}c^2}\xspace}

\def\mum  {\ensuremath{\,\upmu\nospaceunit{m}}\xspace}

\def\fb   {\ensuremath{\aunit{fb}}\xspace}
\def\invfb   {\ensuremath{\fb^{-1}}\xspace}

\def\ps   {\ensuremath{\aunit{ps}}\xspace}

\newcommand{\chisq}{\ensuremath{\chi^2}\xspace}

\newcommand{\chisqip}{\ensuremath{\chi^2_{\text{IP}}}\xspace}

\def\gsim{{~\raise.15em\hbox{$>$}\kern-.85em
          \lower.35em\hbox{$\sim$}~}\xspace}
\def\lsim{{~\raise.15em\hbox{$<$}\kern-.85em
          \lower.35em\hbox{$\sim$}~}\xspace}

\def\pt         {\ensuremath{p_{\mathrm{T}}}\xspace}

\def\ptot       {\ensuremath{p}\xspace}

\def\mrad{\aunit{mrad}}

\def\evtgen     {\mbox{\textsc{EvtGen}}\xspace}

\def\geant      {\mbox{\textsc{Geant4}}\xspace}

\def\photos     {\mbox{\textsc{Photos}}\xspace}

\def\pythia     {\mbox{\textsc{Pythia}}\xspace}

\xspace

\def\tell1  {TELL1\xspace}
\def\ukl1   {UKL1\xspace}

\usepackage{cite} 
\usepackage{mciteplus}
\usepackage{float}

\def\Lcp     {\ensuremath{\Lambdares_\cquark^+}\xspace}

\def\Xicc              {\ensuremath{\Xires_{\cquark\cquark}^+}\xspace}
\def\Xiccpp            {\ensuremath{\Xires_{\cquark\cquark}^{++}}\xspace}
\def\Xiccp             {\Xicc}
\def\Xic               {\ensuremath{\Xires_{\cquark}^{+}}\xspace}

\def\XiccppDecayDp{{\ensuremath{\ensuremath{\Xires_{\cquark\cquark}^{++}}\to\Dp \proton \Km\pip}}\xspace}
\def\XiccppDecayLc{{\ensuremath{\ensuremath{\Xires_{\cquark\cquark}^{++}}\to\Lcp\Km\pip\pip}}\xspace}
\def\XiccppDecayXic{{\ensuremath{\Xiccpp\to\Xic\pip}}\xspace}

\def\DpDecay{{\ensuremath{\Dp\to\Km\pip\pip}}\xspace}

\def\genxicc      {\mbox{\textsc{Genxicc}}\xspace}

\usepackage[mathscr]{euscript} 

\usepackage{longtable} 

\begin{document}

\renewcommand{\thefootnote}{\fnsymbol{footnote}}
\setcounter{footnote}{1}


\begin{titlepage}
\pagenumbering{roman}

\vspace*{-1.5cm}
\centerline{\large EUROPEAN ORGANIZATION FOR NUCLEAR RESEARCH (CERN)}
\vspace*{1.5cm}
\noindent
\begin{tabular*}{\linewidth}{lc@{\extracolsep{\fill}}r@{\extracolsep{0pt}}}
\ifthenelse{\boolean{pdflatex}}
{\vspace*{-1.5cm}\mbox{\!\!\!\includegraphics[width=.14\textwidth]{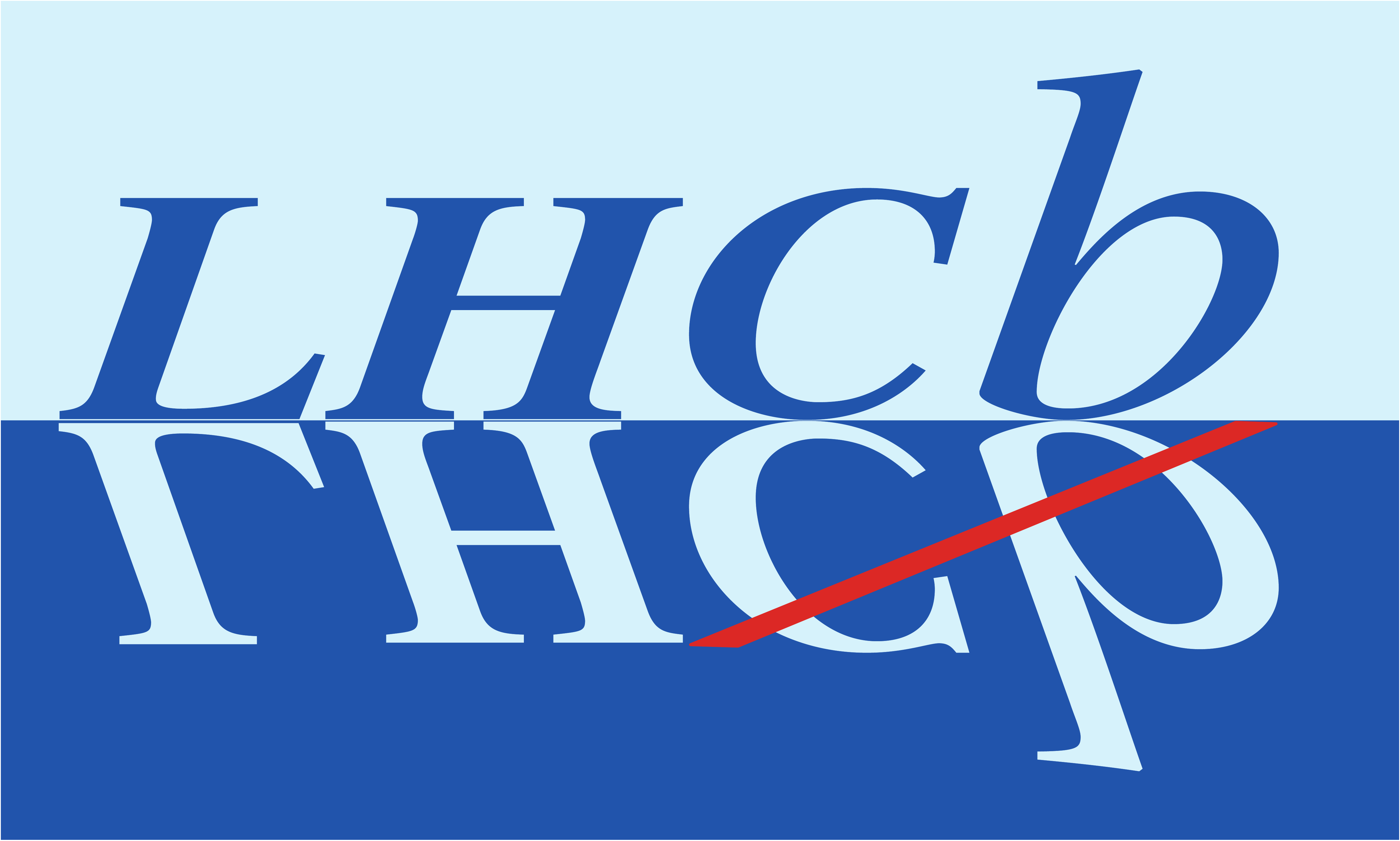}} & &}%
{\vspace*{-1.2cm}\mbox{\!\!\!\includegraphics[width=.12\textwidth]{lhcb-logo.eps}} & &}%
\\
 & & CERN-EP-2019-067 \\  
 & & LHCb-PAPER-2019-011 \\  
 & & October 10, 2019 \\ 
 & & \\
\end{tabular*}

\vspace*{4.0cm}

{\normalfont\bfseries\boldmath\huge
\begin{center}
  \papertitle 
\end{center}
}

\vspace*{2.0cm}

\begin{center}
\paperauthors\footnote{Authors are listed at the end of this paper.}
\end{center}

\vspace{\fill}

\begin{abstract}
  \noindent
  A search for the \Xiccpp~baryon through the \XiccppDecayDp decay is performed 
  with a data sample corresponding to an integrated luminosity of 1.7\invfb 
  recorded by the LHCb experiment in $\proton\proton$ collisions at a centre-of-mass 
  energy of 13\tev. No significant signal is observed in the mass range from the 
  kinematic threshold of the decay to 3800\mevcc. An upper limit is set on the ratio 
  of branching fractions
  $\mathcal{R} =  \frac{\mathcal{B}(\Xiccpp \to \Dp \proton \Km \pip)}{\mathcal{B}
  (\Xiccpp \to \Lcp \Km \pip \pip)} $ with 
  $\mathcal{R} < 1.7 \hspace{2pt} (2.1) \times 10^{-2}$ 
  at the \mbox{90\% (95\%) confidence} level at the known mass of the \Xiccpp~state.
\end{abstract}

\vspace*{2.0cm}

\begin{center}
Published in JHEP, DOI: 10.1007/JHEP10(2019)124
\end{center}

\vspace{\fill}

{\footnotesize 
\centerline{\copyright~\papercopyright. \href{\paperlicenceurl}{\paperlicence}.}}
\vspace*{2mm}

\end{titlepage}


\newpage
\setcounter{page}{2}
\mbox{~}
%
%
%
%

\cleardoublepage


\renewcommand{\thefootnote}{\arabic{footnote}}
\setcounter{footnote}{0}



\pagestyle{plain} 
\setcounter{page}{1}
\pagenumbering{arabic}


\section{Introduction}
\label{sec:intro}

The first observed doubly charged and doubly charmed baryon 
was the \Xiccpp~($ccu$)~state
found through the 
${\XiccppDecayLc}$
and 
${\XiccppDecayXic}$~decay modes
by the LHCb~collaboration~\cite{LHCb-PAPER-2017-018, LHCb-PAPER-2018-026}. 
With two heavy constituent quarks,
this baryon provides a unique system for testing quantum
chromodynamics.
The average mass of the \Xiccpp baryon from the two LHCb measurements 
now stands at
${3621.24 \pm 0.65 (\mathrm{stat}) \pm~0.31~(\mathrm{syst})}$\mevcc
and its lifetime is $0.256^{+0.024}_{-0.022}$~(stat)~$\pm$~0.014~(syst)\ps~\cite{LHCb-PAPER-2018-019}, 
consistent with a weakly decaying state. 
However, many 
features of the \Xiccpp~baryon remain unknown, including its spin and parity.
Previously, signals of the singly charged 
\Xiccp~state were 
reported in the $\Lcp\Km\pip$ and 
$\proton\Dp\Km$~final states by the SELEX collaboration~\cite{Mattson:2002vu,Ocherashvili:2004hi}.
The masses of the \Xiccpp and \Xiccp ground states 
are expected to be approximately equal according to isospin symmetry~\cite{Karliner:2017gml}.
Searches in different production environments at the FOCUS, BaBar, Belle and LHCb experiments 
have however not
shown evidence for a \Xiccp state with the properties reported by
the SELEX collaboration~\cite{ratti2003new,Aubert:2006qw,Chistov:2006zj,LHCb-PAPER-2013-049}.

\begin{figure}
  \begin{center}
    \includegraphics[width = 0.49\linewidth]{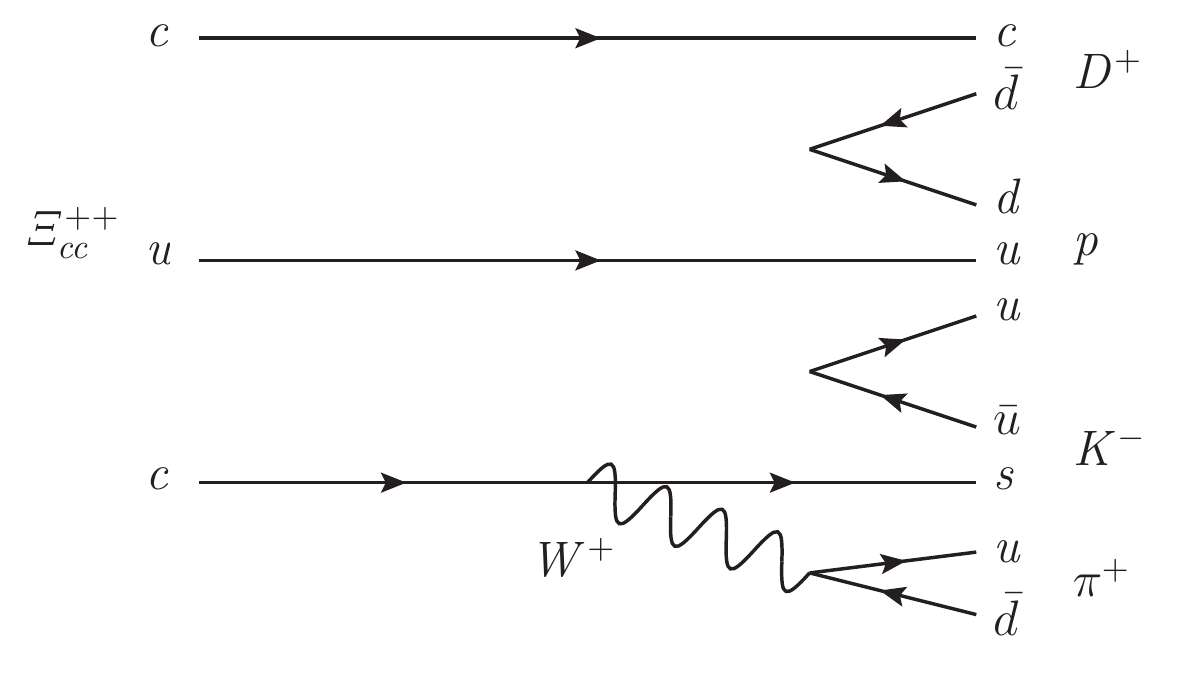}
    \includegraphics[width = 0.49\linewidth]{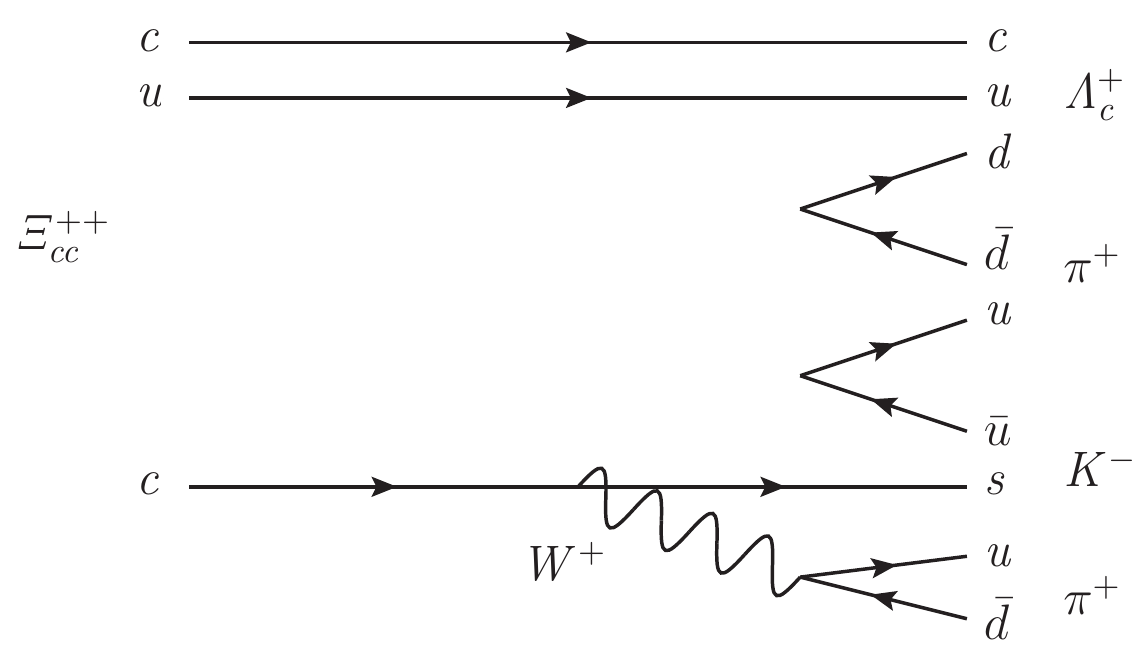}
  \end{center}
  \caption{
    \small
      The Feynman diagram contributing to the inclusive (left) \XiccppDecayDp decay
     with the analogous (right) \XiccppDecayLc diagram.
  }
  \label{fig:FeynmanDiagram}
\end{figure}

To further understand the dynamics of weakly decaying doubly heavy baryons, 
it is of prime importance to pursue searches for additional decay modes of the \Xiccpp baryon.
These decays may 
differ significantly from those of singly heavy hadrons due to interference effects between decay
amplitudes of the two heavy quarks. 
From an experimental viewpoint, 
the decay 
${\Xiccpp \rightarrow \Dp (\rightarrow \Km \pip \pip) \proton \Km \pip}$
is a suitable search channel,
since the ${\DpDecay}$ trigger is 
proven to be very efficient at LHCb~\cite{LHCb-DP-2019-001}.\footnote{The inclusion of charge-conjugate processes is implied throughout this paper.}
The tree-level
amplitudes
of the 
inclusive decays of 
${\XiccppDecayDp}$ and ${\XiccppDecayLc}$,
as shown in Fig.~\ref{fig:FeynmanDiagram}, 
are comparable, which suggests
that the branching fractions of these two modes could be similar. 
Theoretical calculations have been performed on pseudo-two-body
decays of doubly-charmed baryons~\cite{Yu:2017zst}. 
The $\XiccppDecayDp$~decay could 
proceed as a pseudo-two-body decay if it decays via an excited $\Sigmares^{+*}$~state
with a mass greater than 1572\mevcc, which would then decay to a $\proton\Km\pip$~final state.
However, the properties of such $\Sigmares^{+*}$~decays are not well known~\cite{pdg18}. 
The \XiccppDecayDp~decay also has a energy release of 180\mev, 
compared to 560\mev for the ${\XiccppDecayLc}$ decay,
which means it is expected to have a lower branching fraction
because of the smaller available \mbox{phase space.}

The analysis presented in this paper
searches for the
\Xiccpp~baryon, at its known mass, 
through
${\XiccppDecayDp}$ decays
and also explores a larger mass range to identify 
the hypothetical isospin partner of the \Xiccp~state 
that the SELEX collaboration reported.
The analysis uses $\proton\proton$ collision data
corresponding 
to an integrated luminosity of $1.7\invfb$ 
recorded by the LHCb experiment 
in 2016 at a centre-of-mass energy of 13\tev.
The branching fraction of the \XiccppDecayDp decay
is normalised to \XiccppDecayLc
to reduce systematic uncertainties. 

The ratio of branching fractions, $\mathcal{R}$, is determined as
\begin{eqnarray}
\mathcal{R} &=&  \frac{\mathcal{B}(\Xiccpp \to \Dp \proton \Km \pip)}{\mathcal{B}(\Xiccpp \to \Lcp \Km \pip \pip)}  \nonumber \\[0.2cm]
&=&  \frac{\mathcal{B}(\Xiccpp \to \Dp (\to \Km \pip \pip) \proton  \Km \pip)}{\mathcal{B}(\Xiccpp \to \Lcp (\to \proton \Km \pip) \Km \pip \pip)} \times \frac{\mathcal{B} ( \Lcp \to \proton \Km \pip)}{\mathcal{B} (\Dp \to \Km \pip \pip)} \nonumber \\[0.2cm]
&=& \frac{ N(\Dp \proton \Km \pip)}{ N(\Lcp \Km \pip \pip) } \times \frac{ \varepsilon(\Lcp \Km \pip \pip)}{ \varepsilon(\Dp \proton \Km \pip)} 
\times \frac{\mathcal{B} (\Lcp \to \proton \Km \pip)}{\mathcal{B} (\Dp \to \Km \pip \pip) } ,
\label{br_equation}  
\end{eqnarray}
where 
$N(\Dp \proton \Km \pip)$ 
and $N(\Lcp \Km \pip \pip)$ 
refer to the measured yields of the 
signal in the \XiccppDecayDp and \XiccppDecayLc channels, respectively,
and
$\varepsilon(\Dp \proton \Km \pip)$ and $\varepsilon(\Lcp \Km \pip \pip)$
are the corresponding selection efficiencies of the decay modes.
The values for $\mathcal{B} (\Dp \to \Km \pip \pip)$ and $\mathcal{B} (\Lcp \to \proton \Km \pip)$
are known to be ${(8.98 \pm 0.28)\%}$ 
and ${(6.23 \pm 0.33)\%}$, respectively~\cite{pdg18}
and are uncorrelated.

For convenience, 
the single-event sensitivity, $\alpha_{s}$, is defined as
\begin{equation}
  \alpha_{s} \equiv \frac{
   \varepsilon(\Lcp \Km \pip \pip)
  }{
    N(\Lcp \Km \pip \pip) \,  \varepsilon(\Dp p \Km \pip)
  }
  \times \frac{\mathcal{B}( \Lcp \to \proton \Km \pip)}{\mathcal{B} (\Dp \to \Km \pip \pip) } 
  \label{eq:defAlpha}
\end{equation}
such that Eq.~\ref{br_equation} reduces to
$\mathcal{R} = \alpha_{s} \times N(\Dp p \Km \pip)$. 
All aspects of the analysis are fixed before the data
in the  
$[3300,3800]\mevcc$
mass region are examined.


\section{Detector and software}
\label{sec:Detector}

The \lhcb detector~\cite{Alves:2008zz,LHCb-DP-2014-002} is a single-arm forward
spectrometer covering the \mbox{pseudorapidity} range $2<\eta <5$,
designed for the study of particles containing \bquark or \cquark
quarks. The detector includes a high-precision tracking system
consisting of a silicon-strip vertex detector surrounding the $pp$
interaction region~\cite{LHCb-DP-2014-001}, a large-area silicon-strip detector located
upstream of a dipole magnet with a bending power of about
$4{\mathrm{\,Tm}}$, and three stations of silicon-strip detectors and straw
drift tubes~\cite{LHCb-DP-2017-001} placed downstream of the magnet.
The tracking system provides a measurement of the momentum, \ptot, of charged particles with
a relative uncertainty that varies from 0.5\% at low momentum to 1.0\% at 200\gevc.
The minimum distance of a track to a primary vertex (PV), the impact parameter (IP), 
is measured with a resolution of $(15+29/\pt)\mum$,
where \pt is the component of the momentum transverse to the beam, in\,\gevc.
Charged hadrons are identified
using two ring-imaging Cherenkov detectors~\cite{LHCb-DP-2012-003}. 
Photon, electron and
hadron candidates are identified by a calorimeter system consisting of
scintillating-pad, pre-shower detectors, an electromagnetic
calorimeter and a hadronic calorimeter. 
Muons are identified by a
system composed of alternating layers of iron and multiwire
proportional chambers~\cite{LHCb-DP-2012-002}.
The trigger  consists of a
hardware stage, based on information from the calorimeter and muon
systems, followed by a software stage, which applies a full event
reconstruction.
The online reconstruction
incorporates near-real-time alignment and calibration of the detector~\cite{LHCb-DP-2019-001}. 
The same alignment and calibration information is
propagated to the offline reconstruction, ensuring consistent and high-quality 
information 
between the trigger and offline software. 
The identical
performance of the online and offline reconstruction offers the opportunity to perform
physics analyses directly using candidates reconstructed in the trigger~\cite{LHCb-DP-2016-001}.
The analysis described in this paper makes use of these features.

Simulated \XiccppDecayDp decays
are used to design the candidate selection 
and to calculate the efficiency of such a selection.
The proton-proton interactions are generated using
\pythia\cite{Sjostrand:2007gs,*Sjostrand:2006za} 
with a specific \lhcb
configuration~\cite{LHCb-PROC-2010-056}.
\genxicc~v2.0~\cite{Chang:2009va},
the dedicated generator
for doubly-heavy-baryon production at LHCb, is used to produce the signal.
Decays of hadronic particles
are described by \evtgen~\cite{Lange:2001uf}, in which final-state
radiation is generated using \photos~\cite{Golonka:2005pn}. The
interaction of the generated particles with the detector 
and their response 
are implemented using the \geant~toolkit~\cite{Allison:2006ve, *Agostinelli:2002hh} as described in
Ref.~\cite{LHCb-PROC-2011-006}.
The \XiccppDecayDp decays are generated
with a \Xiccpp mass of 3621.40\mevcc
and the 
decay products of 
\Xiccpp and \Dp hadrons 
are distributed
uniformly in phase~space.


\section{Triggering, reconstruction and selection}
\label{sec:sel}

The procedure to trigger, reconstruct
and select candidates is designed to retain \Xiccpp~signal
and to suppress three primary sources of background:
combinatorial background, which
arises from random combination of tracks;
misreconstructed charm or beauty hadron decays, 
which typically have displaced decay vertices;
and 
combinations of a real \Dp~meson 
with other tracks to form a fake \Xiccpp candidate. 
To better control systematic uncertainties,
the selection of \XiccppDecayDp decays is also designed to be 
as similar as possible 
to that of the \XiccppDecayLc normalisation channel,
described in Ref.~\cite{LHCb-PAPER-2017-018}.

The \Dp candidates are reconstructed
in the final state $\Km\pip\pip$.
At least one of the three
tracks used to reconstruct the \Dp~candidate 
must be selected by the inclusive
software trigger, which requires that the track has
\pt $>250\mevc$ and $\chisqip > 4$ with respect to any
PV, where \chisqip is defined as the
difference in \chisq of a given PV reconstructed with and
without the considered track.
The \Dp~candidate then must be reconstructed
and accepted by a dedicated \DpDecay selection
algorithm in the software trigger. 
This algorithm applies several
geometric and kinematic requirements; 
at least one of the three tracks must have ${\pt > 1\gevc}$ and ${\chisqip > 50}$,
at least two of the tracks must have  ${\pt > 0.4\gevc}$ and ${\chisqip > 10}$
and the scalar sum of the \pt of the three tracks
must be larger than 3\gevc.
Furthermore, 
the $\Dp$~candidate must have a good vertex-fit quality with $\chisq/\rm{ndf}<6$.
The candidate 
must also point back to its associated PV, 
where the angle between its flight path 
and momentum vector should be less than 0.01~radians. 
The associated PV 
is that which best fits the flight direction of the reconstructed candidate.
The \Dp~vertex must also be displaced from this PV such that the
estimated \Dp~decay time is longer than $0.4\ps$.
Only candidates whose invariant mass is within $\pm80\mevcc$ of
the known mass of the \Dp~meson (1869.65\mevcc~\cite{pdg18}) are retained. 
Finally, candidates are required to pass a 
MatrixNet classifier~\cite{LHCb-DP-2019-001}
within the software trigger,
which has been trained on $\pt$ and vertex $\chi^{2}$ information
prior to data taking.
For events that pass the online trigger,
the offline selection of \Dp~candidates proceeds in a similar fashion to
that used in the software trigger: three tracks are
required to form a common vertex that is significantly
displaced from 
the associated PV of the candidate
and its combined invariant mass must be in the 
range 
$[1847,1891]\mevcc$.
Particle identification (PID) requirements are imposed on all three
tracks to suppress combinatorial background and
misidentified charm decays. 
The \Xiccpp candidates are formed by combining a \Dp
candidate with three more charged tracks, each with ${\pt>500\mevc}$
and separately identified
as a proton, kaon and pion 
with good track quality.
The three tracks and the \Dp~candidate are required to form a vertex in which each pairwise
combination of the four particles is required to have a distance of closest approach
of less than 10~mm and the fitted \Xiccpp~vertex must have $\chisq/\rm{ndf}<10$.
The \Xiccpp~candidate is also required to point back to the PV, 
and to have $\pt > 4.5$\gevc.
Only events that passed 
the hardware trigger based 
on information from the muon 
and calorimeter systems that are not part of the reconstructed \Xiccpp~event
are used in the analysis~\cite{LHCb-DP-2019-001}.
Hence, the event is triggered independently of the reconstructed \Xiccpp~candidate, 
which reduces the systematic uncertainty on the efficiency ratios
between the \XiccppDecayDp and  \XiccppDecayLc~decay modes. 

To improve the mass resolution, 
the following mass estimator is used in the analysis
\begin{equation}
m(\Dp\proton\Km\pip) \equiv M(\Dp\proton\Km\pip) - M([\Km\pip\pip]_{\Dp}) + M_{\rm{PDG}}(\Dp)
,
\label{eq:defineDeltaM}
\end{equation}
where $M(\Dp\proton\Km\pip)$ is the measured invariant mass of the \Xiccpp candidate,
$M([\Km\pip\pip]_{\Dp})$ is the measured invariant mass of the $\Km \pip \pip$ combination 
corresponding to the intermediate \Dp candidate
and
$M_{\rm{PDG}}(\Dp)$ is the known mass of the \Dp~meson.
By using the mass definition in Eq.~\ref{eq:defineDeltaM}, 
a mild correlation between decay~time 
and mass is reduced and the mass resolution is improved by 0.15\mevcc.
The \Xiccpp~candidates are 
accepted if they have a reconstructed mass in the range 
${3300 \leq m(\Dp\proton\Km\pip) \leq 3800}$\mevcc.

Following a comparison study of different multivariate methods,
a classifier based on the multilayer perceptron (MLP) algorithm~\cite{Hocker:2007ht}
is used to further suppress combinatorial background.
Simulated \Xiccpp decays 
are used
to train the MLP classifier to recognise signal.
Dedicated software triggers reconstruct 
an unphysical combination of $\Dp \proton \Kp \pip$ (wrong-sign-plus, WSP) 
and $\Dp \proton \Km \pim$ (wrong-sign-minus, WSM) data.
The WSP and WSM samples
are expected to be good 
proxies for combinatorial background in the \XiccppDecayDp (right-sign, RS) channel.
For this analysis, WSP data in the 
${3550 \leq m(\Dp\proton\Km\pip) \leq 3700\mevcc}$ mass region
is used to train the MLP~classifier to 
identify background,
while the WSM data is used to cross-check the results.
Fifteen input variables are used in the MLP training.
The variables with the best discriminating power between signal and background are:
the \Xiccpp vertex fit with a kinematic refit~\cite{Hulsbergen:2005pu} of the \Xiccpp decay chain requiring it to originate from its PV; 
the smallest \pt of the four decay products of the \Xiccpp candidate; 
the angle between the \Xiccpp~momentum vector 
and the direction from the PV to the \Xiccpp decay vertex;
the $\chisqip$ of the \Xiccpp candidate with respect to its PV;
the maximum distance of the closest approach between all pairs of \Xiccpp~tracks forming the \Xiccpp candidate; 
and the maximum distance of the closest approach 
between 
all pairs of tracks from the decay of the \Dp~candidate.
To maintain a sizeable number of signal events, 
the hardware-trigger requirements
are not applied to the signal and background samples. 
In addition to the training samples, 
disjoint testing samples are acquired from the same source.
After training, the response of the MLP is compared
between the training and testing samples. 
No signs of the MLP~classifier being overtrained 
are found based on 
the Kolmogorov--Smirnov test statistic.
Candidates are retained only if the MLP response output 
exceeds a certain threshold. 
The threshold is chosen by maximising the Punzi figure of merit~\cite{Punzi:2003bu},
with a target significance of five sigma.
To test for potential misreconstruction effects,
the same selection criteria are applied to the WSP and WSM data;
no peaking structures are visible in either control sample, as expected.

After the multivariate selection, events may
contain multiple \Xiccpp candidates.
This can arise from mistakes in the reconstruction
of
${\Xiccpp \rightarrow \Dp (\rightarrow \Km \pip \pip) \proton \Km \pip}$
decays. 
For instance, 
there can be cases 
when \Xiccpp~candidates in the same event
have used the same track more than once.
To deal with this, 
the angle between any two tracks of the same charge 
is required to be greater than 0.5\mrad.
If a \Xiccpp~candidate has been formed from 
at least one pair of these cloned tracks, then the candidate is removed.  
This requirement removes 
around 6\% of \Xiccpp~candidates in RS data following the multivariate selection.
In a separate scenario,
the same six final-state tracks may be used to reconstruct
more than one \Xiccpp~candidate in the same event but with 
the tracks wrongly interchanged
(e.g., the \Km~track originating from the \Xiccpp~decay vertex 
and the \Km~track coming from the \Dp~decay vertex). 
In this situation, 
only one of the \Xiccpp~candidate 
from such an event, chosen at random, is retained.
This requirement discards less than 1\% of candidates
at this stage of the selection. 


\section{Mass distributions}
\label{sec:mass}

To determine the yield of \Xiccpp and \Dp particles following the selection
of ${\XiccppDecayDp}$ candidates, the $m(\Dp\proton\Km\pip)$ and
$M([\Km\pip\pip]_{\Dp})$ mass distributions are fitted using models that are
developed using simulation.

\begin{figure}[htp]
  \begin{center}
  \includegraphics[width = 0.49\linewidth]{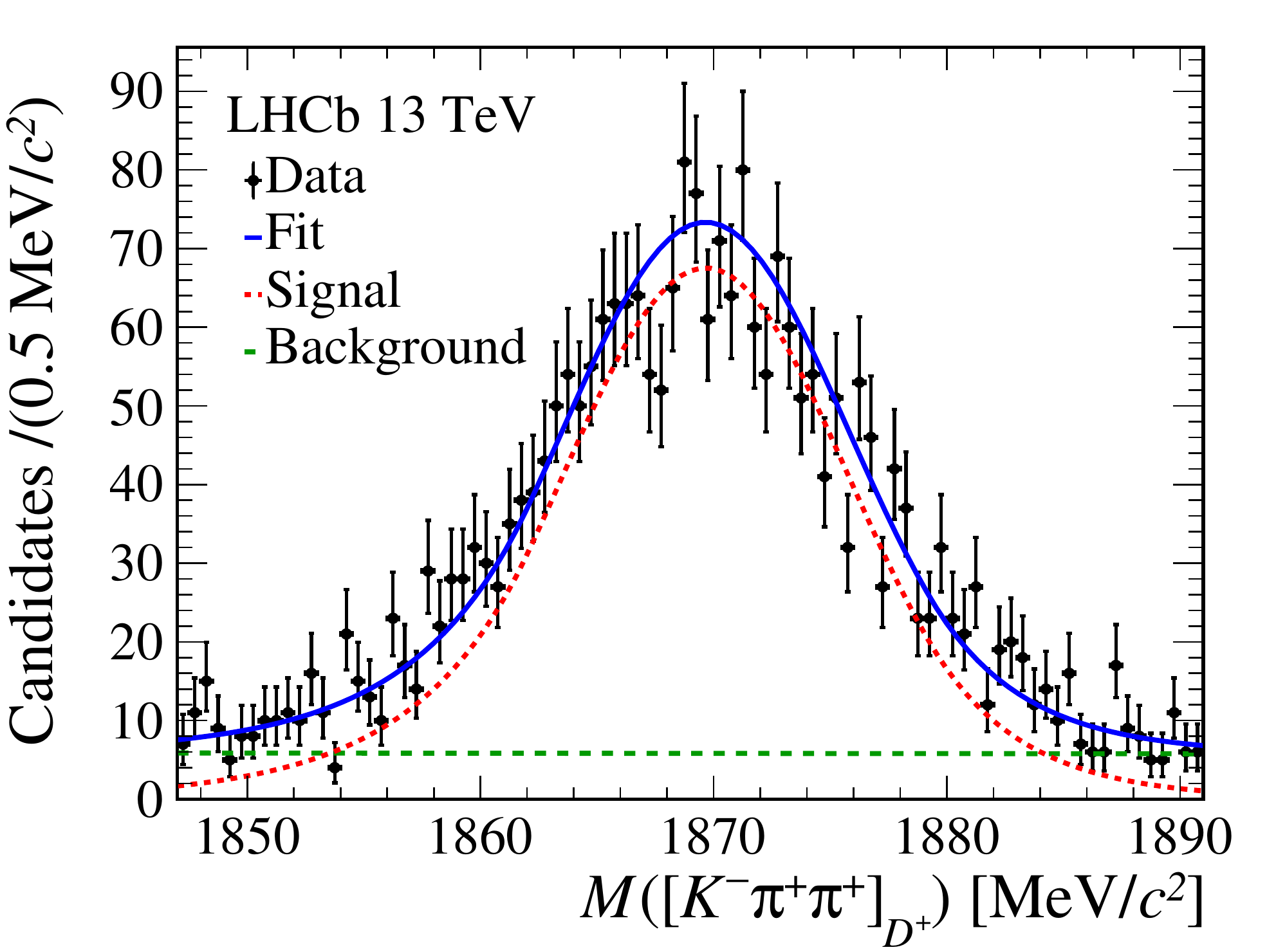}
  \includegraphics[width = 0.49\linewidth]{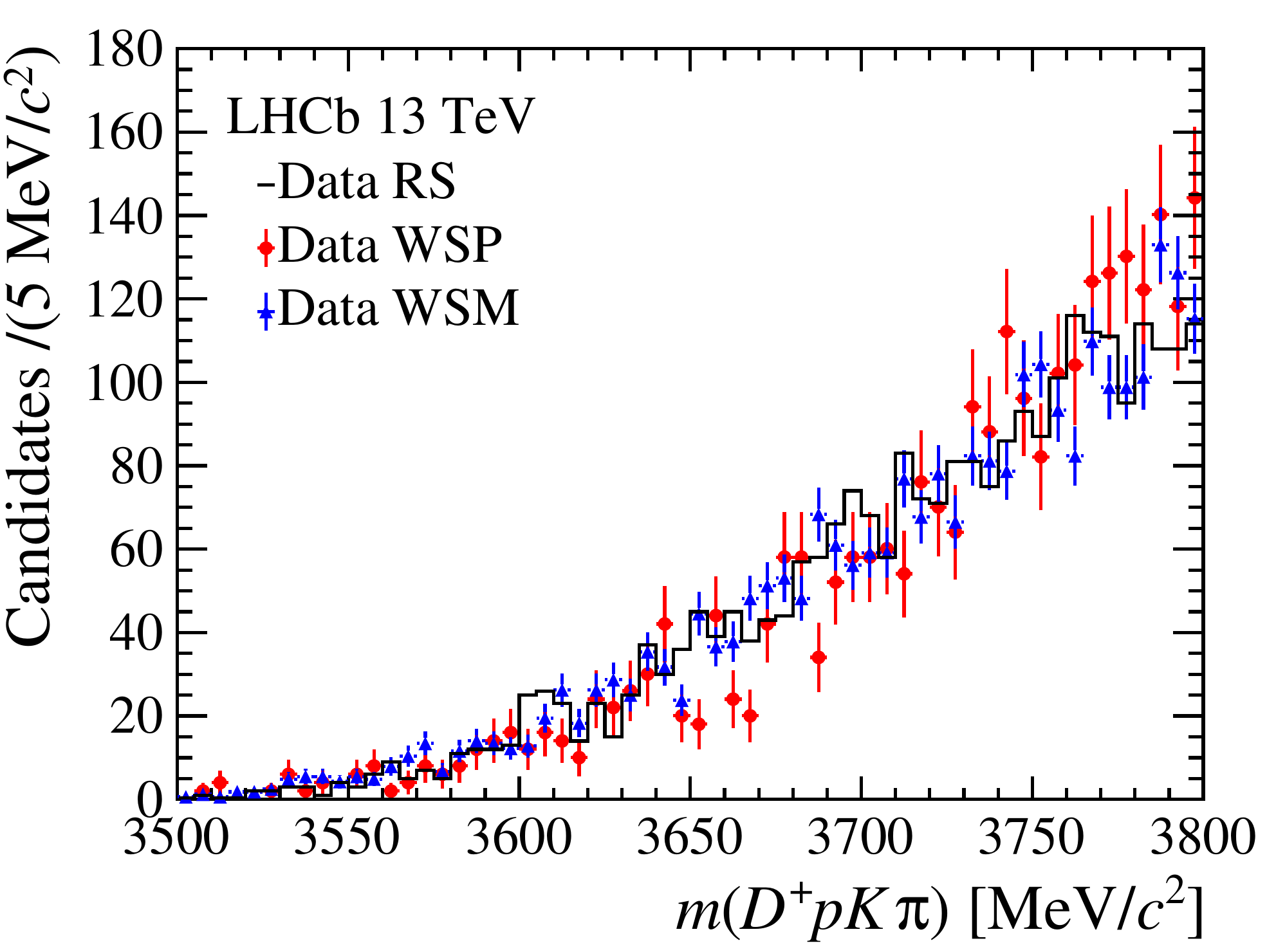}
    \end{center}
  \caption{
    \small
  (left) Invariant-mass distribution of the \Dp candidates after the full selection. 
  The black points represent data and the fit is indicated by the continuous (blue) 
  line with the individual signal and background components represented by the dotted
  (red) line and dashed (green) line, respectively.
  (right) Invariant-mass distributions of right-sign (black) $\Dp\proton\Km\pip$,
  wrong-sign-plus (red) $\Dp\proton\Kp\pip$ and wrong-sign-minus (blue) 
  $\Dp\proton\Km\pim$ data combinations are shown. The control samples have been normalised to the right-sign sample. 
  }
  \label{fig:Dp_mass}
\end{figure}

The invariant-mass distribution $M([\Km\pip\pip]_{\Dp})$ of the \Dp~candidates  after the candidate selection is shown in Fig.~\ref{fig:Dp_mass} (left).
A Crystal Ball function with exponential tails on both sides~\cite{Skwarnicki:1986xj}
is used to model the signal component and a linear function is used to fit the 
background contribution. The parameters of the signal model are fixed to values 
obtained from simulation, while all parameters in the background model are free.
The selection retains 2697 \Dp~candidates with a purity of 80\% according to 
the results of the fit to the mass spectrum.

The invariant-mass distributions in the RS, WSP and WSM data samples
after the candidate selection are shown in Fig.~\ref{fig:Dp_mass} (right). All 
the samples have similar smoothly shaped distributions across the entire mass
range studied.

\begin{figure}[htp]
  \begin{center}
    \includegraphics[width = 0.49\linewidth]{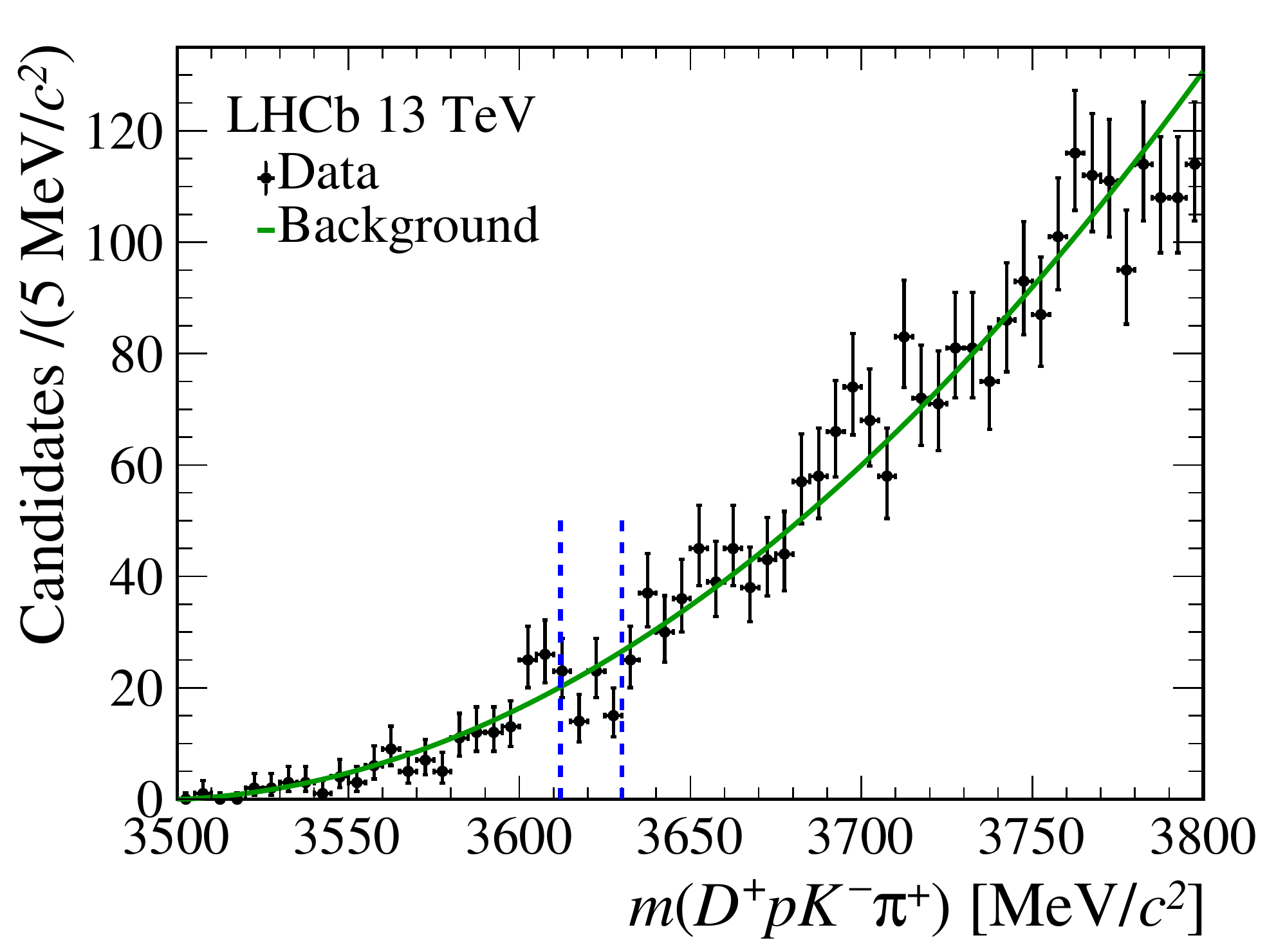}
    \includegraphics[width = 0.49\linewidth]{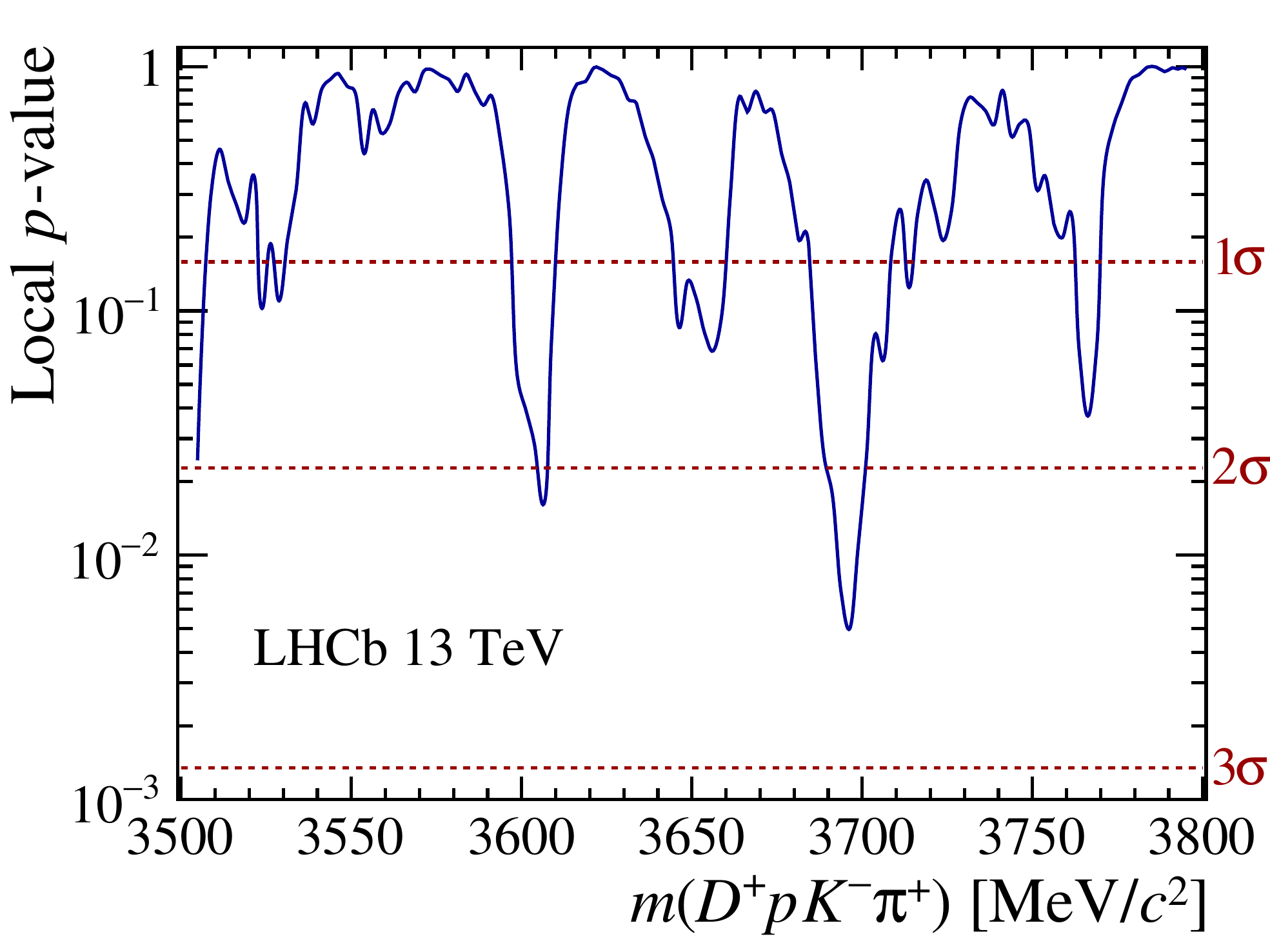}  
  \end{center}
  \caption{
    \small
    (left) Invariant-mass distribution of the ${\XiccppDecayDp}$ candidates
    with the fit overlaid. The black points represent data, the continuous 
    (green) line represents the combinatorial background and the two vertical 
    parallel dashed (blue) lines define the region where the signal is expected.
    (right) The local \emph{p}-value expressing the compatibility of the data 
    with the background-only hypothesis. The horizontal dashed (red) lines indicate \emph{p}-values of 1, 2 and 3$\sigma$ local significance.  
  }
  \label{fig:Xicc_mass}
\end{figure}

The invariant-mass distribution of the \Xiccpp candidates, $m(\Dp\proton\Km\pip)$,
for the signal decay mode after applying all requirements of the analysis, is shown 
in Fig.~\ref{fig:Xicc_mass} (left). The mass distribution is fitted 
with an unbinned extended maximum-likelihood method, assuming only a background
contribution, described by a second-order Chebyshev polynomial. No signal peak is 
visible in the spectrum and the local \emph{p}-value is calculated as a function 
of mass and shown in Fig.~\ref{fig:Xicc_mass} (right). The local 
\emph{p}-value is defined as the probability of observing data that is less 
compatible with the background-only hypothesis than the data set. The test statistic
used is based on $q_0$ in Ref.~\cite{Cowan:2010js}, but instead of assigning it the 
value zero when observing fewer than expected candidates, it is assigned the value $-q_0$ 
to achieve a more intuitive behaviour of the \emph{p}-value for downward 
fluctuations. The likelihoods are evaluated with Poisson statistics using the predicted 
number of background candidates and observed number of signal candidates in regions of
$\pm3\sigma_m$ around each hypothetical mass, where $\sigma_m=2.8\mevcc$ is the
\Xiccpp mass resolution determined from simulated \XiccppDecayDp decays.

There is no visible signal near the mass of 3620\mevcc where a \Xiccpp signal would be
expected, nor is there any excess of candidates near the mass of 3520\mevcc where the
hypothetical isospin partner was observed by the SELEX
collaboration~\cite{Mattson:2002vu,Ocherashvili:2004hi}. The global \emph{p}-value,
including the look-elsewhere effect in the mass range $3500-3800\mevcc$, is 26\% 
and only one signal candidate is observed in the mass range from the kinematic 
threshold of 3441\mevcc to 3500\mevcc. Hence, no significant signal is observed 
in the mass range from the kinematic threshold to 3800\mevcc and we proceed to 
set a limit on the relative branching fraction $\mathcal{R}$.

\begin{figure}[htp]
  \begin{center}
    \includegraphics[width = 0.6\linewidth]{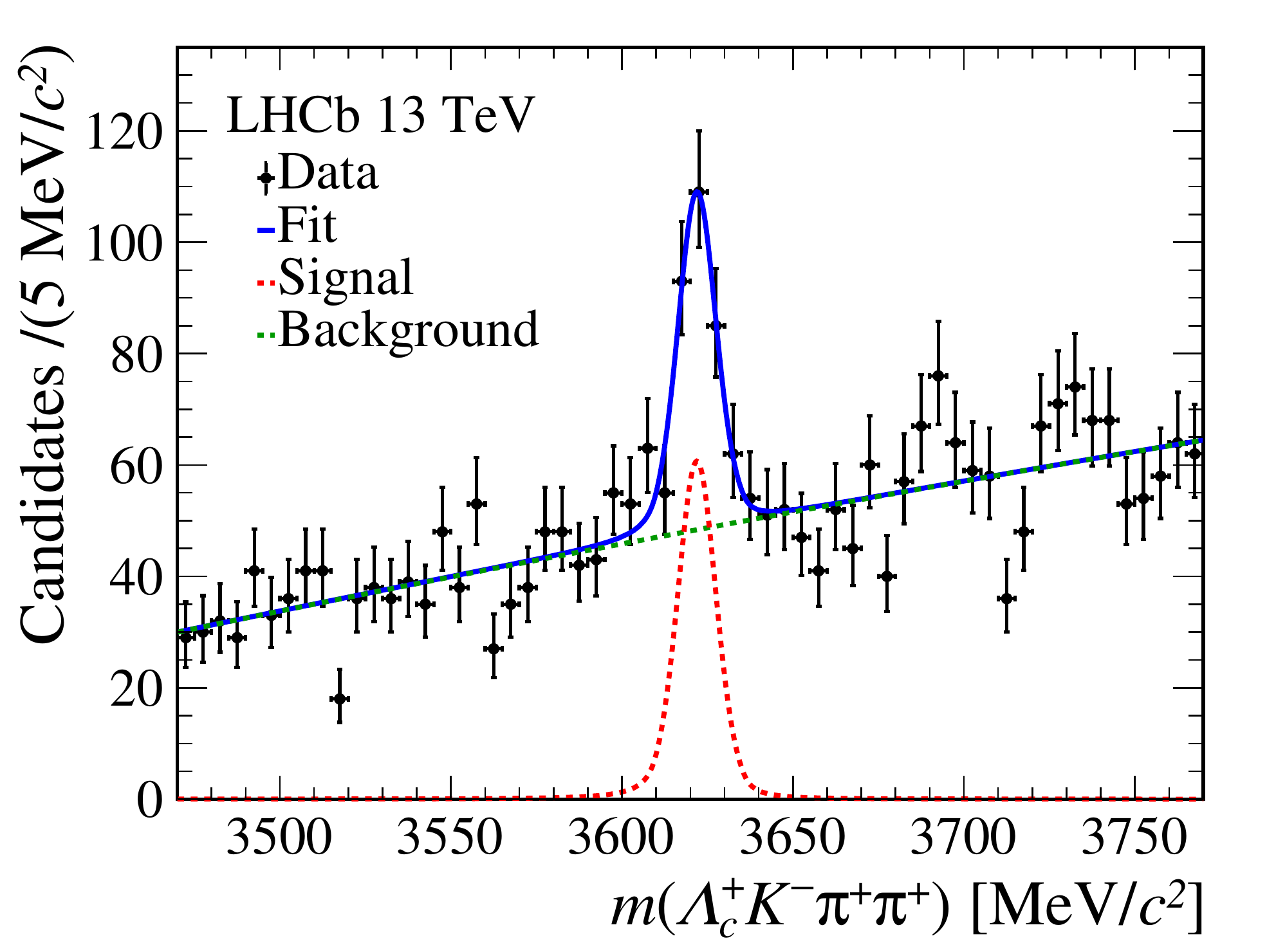}
  \end{center}
  \caption{
    \small
    Invariant-mass distribution of the ${\XiccppDecayLc}$ candidates with the 
    fit overlaid. The black points represent data, the dashed (green) line 
    represents combinatorial background, the dotted (red) line represents 
    the signal contribution and the continuous (blue) line is the total fit.
}
  \label{fig:control_modes_mass}
\end{figure}

The invariant-mass distribution of the \Xiccpp candidates, $m(\Lcp\Km\pip\pip)$, 
for the normalization decay mode, \XiccppDecayLc, is shown in 
Fig.~\ref{fig:control_modes_mass}. In this case a signal peak is clearly visible.
Both the candidate selection and the modelling of the mass spectrum  are identical 
to that in Ref.\cite{LHCb-PAPER-2017-018}, except for the additional requirements 
on the hardware trigger. An extended unbinned maximum-likelihood fit to this 
invariant-mass distribution returns a signal yield of $184\pm29$.

\section{Efficiency determination}
\label{sec:eff}

To set an upper limit on the ratio $\mathcal{R}$, it is necessary to
evaluate
the ratio of efficiencies
between the \XiccppDecayDp and \XiccppDecayLc
decay modes.

The efficiency ratio may be factorised as
\begin{equation}
\label{eq:eff}
  \frac{
	\varepsilon(\Lcp \Km \pip \pip)
  }{
	 \varepsilon(\Dp \proton \Km \pip)
  } = 
  \frac{ \varepsilon_{\Lcp \Km \pip \pip}^{\text{acc}} }{ \varepsilon_{\Dp \proton \Km \pip}^{\text{acc}} } \,
  \frac{ \varepsilon_{\Lcp \Km \pip \pip}^{\text{sel}|\text{acc}} }{ \varepsilon_{\Dp \proton \Km \pip}^{\text{sel}|\text{acc}} } \,
  \frac{ \varepsilon_{\Lcp \Km \pip \pip}^{\text{PID}|\text{sel}} }{ \varepsilon_{\Dp \proton \Km \pip}^{\text{PID}|\text{sel}} } \,
  \frac{ \varepsilon_{\Lcp \Km \pip \pip}^{\text{trig}|\text{PID}} }{ \varepsilon_{\Dp \proton \Km \pip}^{\text{trig}|\text{PID}} }
  ,
\end{equation}
where efficiencies are evaluated for
the geometric acceptance (acc),
the reconstruction and selection excluding particle identification requirements (sel),
the particle identification requirements (PID)
and the trigger (trig).
Each factor is the efficiency relative to all previous steps in the
order given above. 
The individual ratios are evaluated with
simulated \Xiccpp decays,
assuming a uniform phase space model, 
except for PID which is derived from data~\cite{LHCb-PUB-2016-021, LHCb-DP-2012-003}.
The efficiencies are corrected for known differences 
between simulation and data, apart from the geometric acceptance.

The individual efficiency components, shown in Eq.~\ref{eq:eff},
are found to be similar 
between the two \Xiccpp decay modes,
except for
the reconstruction and selection efficiency,
$\varepsilon^{\text{sel}|\text{acc}}$,
where in the \XiccppDecayDp channel
it is found to be approximately twice as large 
as that of the \XiccppDecayLc decay.
This leads
to a total efficiency ratio of 
$\varepsilon(\Lcp \Km \pip \pip) /  \varepsilon(\Dp \proton \Km \pip) = 0.46\pm0.01$, 
where the uncertainty is statistical only.
Combining this total relative efficiency with the value for $N(\Lcp \Km \pip \pip)$
obtained in Sect.~\ref{sec:mass}
and the known values for the branching fractions 
${\mathcal{B} (\Dp \to \Km \pip \pip)}$
and 
${\mathcal{B} (\Lcp \to \proton \Km \pip)}$,
then
according to Eq.~\ref{eq:defAlpha},
the single-event sensitivity is 
${\alpha_{s} =(1.74 \pm 0.29) \times 10^{-3}}$.
The uncertainty on ${\alpha_{s}}$ includes the total uncertainty on
the ${\mathcal{B} (\Dp \to \Km \pip \pip)}$
and ${\mathcal{B} (\Lcp \to \proton \Km \pip)}$ branching fractions
and the statistical uncertainty on the 
$N(\Lcp \Km \pip \pip)$ and 
$\varepsilon(\Lcp \Km \pip \pip) /  \varepsilon(\Dp \proton \Km \pip)$ 
measured values.


\section{Systematic uncertainties}
\label{sec:sys}

The statistical uncertainty on the measured signal yield 
in the \XiccppDecayLc channel
is the dominant uncertainty on $\alpha_{s}$ 
and the systematic uncertainties on $\alpha_{s}$ 
have small effect
on the upper limits on the ratio $\mathcal{R}$.

The largest systematic uncertainty arises from the evaluation of the 
efficiency of the 
hardware-trigger requirement.
Only candidates that 
are triggered independently of the \Xiccpp candidate's final-state tracks
are used in the branching fraction ratio limit to minimise this systematic uncertainty.
The ratio of these efficiencies is equal 
to one if the kinematic distributions of the \Xiccpp~candidate 
in the \XiccppDecayDp and \XiccppDecayLc decay modes are identical.
However, the efficiencies can be different 
if the respective selection requirements 
of the \XiccppDecayDp
and \XiccppDecayLc analyses 
select different kinematic regions of the \Xiccpp~candidate. 
This effect is studied by weighting the \pt distributions 
in simulated samples. 
The change in efficiency of the hardware trigger after the weighting 
is evaluated and results in a systematic uncertainty of 3.5\%.
The impact of the model used to fit the $m(\Lcp\Km\pip\pip)$ invariant-mass distribution
on the yield of \Xiccpp~candidates, $N(\Lcp \Km \pip \pip)$,
is investigated by using alternative signal and background models 
and performing the fit
over different mass ranges.
The largest variation in the yield of \Xiccpp candidates is 3.1\% and this is taken as a systematic uncertainty on $\alpha_{s}$. 
The effect of the uncertainty associated with the \Xiccpp baryon's 
lifetime on 
the relative reconstruction and selection efficiency
between the \XiccppDecayDp and \XiccppDecayLc channels 
is investigated by varying the lifetime within its uncertainty
and a systematic uncertainty of 2.9\% is assigned to the $\alpha_{s}$ parameter.
The PID efficiency is determined in bins of particle momentum 
and pseudorapidity using calibration samples taken from data~\cite{LHCb-PUB-2016-021}.
The size of the bins is increased or decreased by a factor of two
and the largest deviation on $\alpha_{s}$ of 1.5\%
is assigned as systematic uncertainty.
Finally, since the simulation may not describe the signal perfectly, simulated
\XiccppDecayDp decays are weighted to make their \pt distribution match that
observed in the \XiccppDecayLc data. 
The selection and software-trigger efficiencies are similarly calculated using
\mbox{\pt-corrected} simulated \Xiccpp~decays. The number of \pt bins used
is increased or decreased by a factor of two and the efficiencies are recalculated 
for both decay channels. This results in a change in $\alpha_{s}$ of 1.2\%.
All efficiencies calculated from simulation are averaged over the entire phase space 
assuming a uniform distribution for both the ${\XiccppDecayDp}$ and
${\XiccppDecayLc}$ decays.
The phase-space distributions of the selected candidates are uniform and show agreement 
in data and simulation. Therefore, no systematic uncertainty 
is assigned to the relative selection and reconstruction efficiencies for the effect 
of intermediate resonances in their decay.

Table~\ref{tab:sys} summarises the systematic and statistical uncertainties on 
$\alpha_{s}$. The statistical uncertainty is dominated by the uncertainty of the 
yield of the normalisation mode but includes a small contribution from the 
finite size of the simulated samples. The ratio of the branching fractions  
${\mathcal{B} (\Dp \to \Km \pip \pip)}$ and 
${\mathcal{B} (\Lcp \to \proton \Km \pip)}$ have a combined uncertainty of 5.7\%.
The systematic uncertainties from the different sources discussed above are considered
uncorrelated and are added in quadrature to give  a total systematic uncertainty of 
5.8\%. Adding all sources of uncertainty in quadrature gives a total uncertainty 
of 17.7\% on the $\alpha_{s}$ parameter. 

\begin{table}
  \caption{
    \small
    Systematic and statistical uncertainties on the single-event sensitivity $\alpha_{s}$.
  }
  \begin{center}
    \begin{tabular}[c]{lc}
      \hline
      \multicolumn{1}{l}{Source}  & \multicolumn{1}{c}{$\alpha_{s}\ (\%)$} \\
      \hline
      Statistical & $15.7$ \\
      Branching fractions & $\phantom{0}5.7$ \\
      \hline
      Trigger efficiency      &     $\phantom{0}3.5$   \\
      Mass fit model 	 &     $\phantom{0}3.1$   \\
      \Xiccpp lifetime  &     $\phantom{0}2.9$   \\
      PID calibration         &     $\phantom{0}1.5$  \\
      Simulation modelling & $\phantom{0}1.2$  \\
      \hline 
      Total uncertainty       &     $17.7$  \\
      \hline
    \end{tabular}
  \end{center}
  \label{tab:sys}
\end{table}


\section{Results}
\label{sec:results}

In this analysis no significant \XiccppDecayDp signal is observed
so an upper limit is set on the ratio of branching fractions, $\mathcal{R}$.
The \texttt{CLs} method~\cite{Read:2002hq} is used to determine 
the ratio of confidence levels~(CL) between the signal-plus-background
and background-only hypotheses.
The upper limit is obtained from the total number of candidates, 
$N_{\rm{obs}}$, observed in the expected signal mass region. 
This value is calculated by counting the number of candidates within the mass region,
${3612 < m(\Dp\proton\Km\pip) < 3630\mevcc}$
(indicated by two dashed blue lines in the left-hand plot of Fig.~\ref{fig:Xicc_mass}).
This mass region corresponds to approximately $\pm 3 \sigma_m$ around the
average mass of the \Xiccpp  state.

The \texttt{CLs} score for a possible value of ratio~$\mathcal{R}$ is calculated as
\begin{equation}
\texttt{CLs} = \frac{ P(N_{\rm{b}}+N_{\rm{s}} \leq N_{\rm{obs}} ) } { P(N_{\rm{b}} \leq N_{\rm{obs}} )},
\label{eqn:CL_toy}
\end{equation}
where $N_{\rm{s}}$ is sampled from the distribution of the expected number of signal
candidates for a given ratio $\mathcal{R}$, $N_{\rm{b}}$ is sampled from the distribution 
of the expected number of background candidates predicted by the background-only fit
(Fig.~\ref{fig:Xicc_mass}, left) and $P$~indicates the probability that these 
statistical quantities are smaller than $N_{\rm{obs}}$.
The data points in the mass region ${3612 < m(\Dp\proton\Km\pip) < 3630\mevcc}$ 
are removed for the fit and $N_{\rm{b}}$ is determined by performing an integral
extrapolation. 
The probability requirements in the numerator and denominator of Eq.~\ref{eqn:CL_toy}
are tested by running a large number of pseudoexperiments 
sampling from a Poisson distribution 
with statistical means of $N_{\rm{b}} + N_{\rm{s}}$ and $N_{\rm{b}}$, respectively. 
The 17.7\% uncertainty on $\alpha_{s}$ is fully accounted 
for by sampling from a Gaussian distribution in each pseudoexperiment. 

\begin{figure}
  \begin{center}
    \includegraphics[width = 0.6\linewidth]{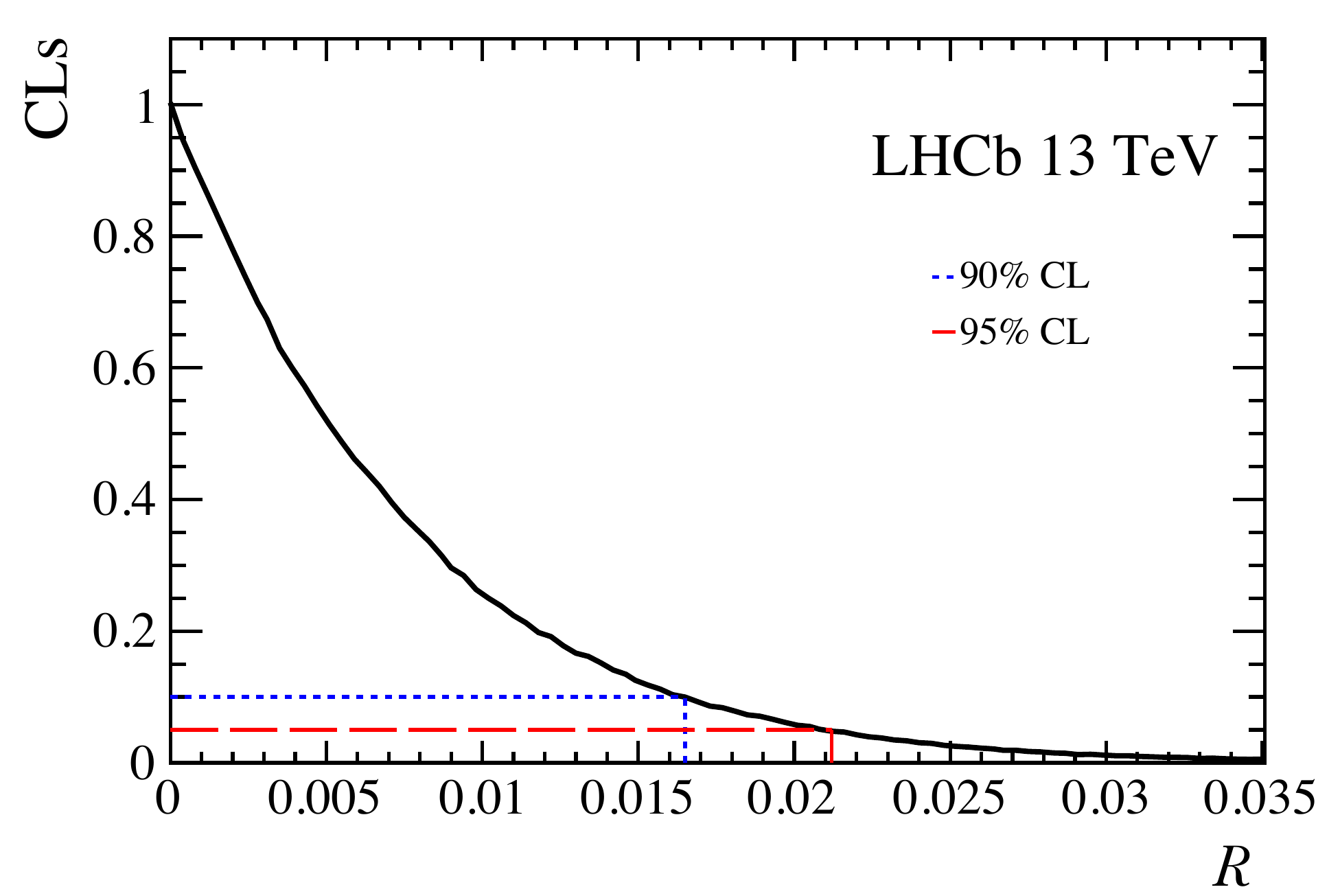}
  \end{center}
  \caption{
    \small
    The scores from the \texttt{CLs} method for each value of the assumed ratio of branching fractions~$\mathcal{R}$. 
    Observed values are shown by the solid black line. 
    The set upper limits at 90\% and 95\% CL are indicated by 
    the dotted (blue) line 
    and
    the dashed (red) line, 
    respectively.
  }
  \label{fig:cls}
\end{figure}

The derived \texttt{CLs} curve as a function of the possible values of the ratio~$\mathcal{R}$ 
is shown as the black line in Fig.~\ref{fig:cls}.
This curve is obtained using values of $N_{\rm obs} = 66$ and $N_{\rm b} = 79.8$ 
as observables 
and running $1 \times 10^{6}$ pseudoexperiments for each hypothetical value of ratio~$\mathcal{R}$. 
The upper limit measured is
\begin{equation}
\mathcal{R} < 1.7 \hspace{3pt}(2.1) \times 10^{-2} \hspace{5pt} \rm{at} \hspace{5pt} 90\% \hspace{3pt}  (95\%) \hspace{5pt} \text{CL}  \nonumber
\end{equation}
as shown by the blue dotted line (red dashed line) in Fig.~\ref{fig:cls}.

\section{Conclusions}
\label{sec:conc}

Following observations of the \XiccppDecayLc and \XiccppDecayXic decay modes, 
a search for the decay \XiccppDecayDp  is performed using $\proton\proton$ 
collision data recorded by the LHCb experiment in 2016 
at a centre-of-mass energy of 13\tev, corresponding to an integrated luminosity 
of $1.7\invfb$. No significant signal is found in the mass range from the kinematic
threshold of the decay of 3441\mevcc to 3800\mevcc. Considering the statistical and
systematic uncertainties, an upper limit on the ratio of branching fractions between 
the \XiccppDecayDp and \XiccppDecayLc decay is set to be
$\mathcal{R} < 1.7 \hspace{1pt} (2.1) \times 10^{-2}$ 
at the 90\% (95\%) confidence level at the known mass of the \Xiccpp~baryon.

The upper limit on the ratio of branching fractions between the two \Xiccpp~decay modes
is derived assuming a uniform phase space model in the efficiency determinations. 
A better theoretical understanding of the resonant and nonresonant contributions
underpinning the \XiccppDecayDp and \XiccppDecayLc decay processes is required to 
understand the at least two orders of magnitude difference between the branching 
fractions of the two \Xiccpp decay modes.
Dynamical effects or spin constraints in the resonance structures could be 
suppressing the \XiccppDecayDp decay.
The full dataset from LHCb, or future data taking with the upgraded detector, 
may reveal evidence of this decay and then shed more light on the production 
and decay dynamics of the \Xiccpp~baryon.

\section*{Acknowledgements}
%
%
\noindent We express our gratitude to our colleagues in the CERN
accelerator departments for the excellent performance of the LHC. We
thank the technical and administrative staff at the LHCb
institutes.
We acknowledge support from CERN and from the national agencies:
CAPES, CNPq, FAPERJ and FINEP (Brazil); 
MOST and NSFC (China); 
CNRS/IN2P3 (France); 
BMBF, DFG and MPG (Germany); 
INFN (Italy); 
NWO (Netherlands); 
MNiSW and NCN (Poland); 
MEN/IFA (Romania); 
MSHE (Russia); 
MinECo (Spain); 
SNSF and SER (Switzerland); 
NASU (Ukraine); 
STFC (United Kingdom); 
DOE NP and NSF (USA).
We acknowledge the computing resources that are provided by CERN, IN2P3
(France), KIT and DESY (Germany), INFN (Italy), SURF (Netherlands),
PIC (Spain), GridPP (United Kingdom), RRCKI and Yandex
LLC (Russia), CSCS (Switzerland), IFIN-HH (Romania), CBPF (Brazil),
PL-GRID (Poland) and OSC (USA).
We are indebted to the communities behind the multiple open-source
software packages on which we depend.
Individual groups or members have received support from
AvH Foundation (Germany);
EPLANET, Marie Sk\l{}odowska-Curie Actions and ERC (European Union);
ANR, Labex P2IO and OCEVU, and R\'{e}gion Auvergne-Rh\^{o}ne-Alpes (France);
Key Research Program of Frontier Sciences of CAS, CAS PIFI, and the Thousand Talents Program (China);
RFBR, RSF and Yandex LLC (Russia);
GVA, XuntaGal and GENCAT (Spain);
the Royal Society
and the Leverhulme Trust (United Kingdom).

\addcontentsline{toc}{section}{References}
\bibliographystyle{LHCb}
\bibliography{main,standard,LHCb-PAPER,LHCb-CONF,LHCb-DP,LHCb-TDR,Xicc}

\ifx\mcitethebibliography\mciteundefinedmacro
\PackageError{LHCb.bst}{mciteplus.sty has not been loaded}
{This bibstyle requires the use of the mciteplus package.}\fi
\providecommand{\href}[2]{#2}
\begin{mcitethebibliography}{10}
\mciteSetBstSublistMode{n}
\mciteSetBstMaxWidthForm{subitem}{\alph{mcitesubitemcount})}
\mciteSetBstSublistLabelBeginEnd{\mcitemaxwidthsubitemform\space}
{\relax}{\relax}

\bibitem{LHCb-PAPER-2017-018}
LHCb collaboration, R.~Aaij {\em et~al.},
  \ifthenelse{\boolean{articletitles}}{\emph{{Observation of the doubly charmed
  baryon $\Xires_{cc}^{++}$}},
  }{}\href{https://doi.org/10.1103/PhysRevLett.119.112001}{Phys.\ Rev.\ Lett.\
  \textbf{119} (2017) 112001},
  \href{http://arxiv.org/abs/1707.01621}{{\normalfont\ttfamily
  arXiv:1707.01621}}\relax
\mciteBstWouldAddEndPuncttrue
\mciteSetBstMidEndSepPunct{\mcitedefaultmidpunct}
{\mcitedefaultendpunct}{\mcitedefaultseppunct}\relax
\EndOfBibitem
\bibitem{LHCb-PAPER-2018-026}
LHCb collaboration, R.~Aaij {\em et~al.},
  \ifthenelse{\boolean{articletitles}}{\emph{{First observation of the doubly
  charmed baryon decay \decay{\Xires_{cc}^{++}}{\Xires_c^+\pi^+}}},
  }{}\href{https://doi.org/10.1103/PhysRevLett.121.162002}{Phys.\ Rev.\ Lett.\
  \textbf{121} (2018) 162002},
  \href{http://arxiv.org/abs/1807.01919}{{\normalfont\ttfamily
  arXiv:1807.01919}}\relax
\mciteBstWouldAddEndPuncttrue
\mciteSetBstMidEndSepPunct{\mcitedefaultmidpunct}
{\mcitedefaultendpunct}{\mcitedefaultseppunct}\relax
\EndOfBibitem
\bibitem{LHCb-PAPER-2018-019}
LHCb collaboration, R.~Aaij {\em et~al.},
  \ifthenelse{\boolean{articletitles}}{\emph{{Measurement of the lifetime of
  the doubly charmed baryon $\Xires_{cc}^{++}$}},
  }{}\href{https://doi.org/10.1103/PhysRevLett.121.052002}{Phys.\ Rev.\ Lett.\
  \textbf{121} (2018) 052002},
  \href{http://arxiv.org/abs/1806.02744}{{\normalfont\ttfamily
  arXiv:1806.02744}}\relax
\mciteBstWouldAddEndPuncttrue
\mciteSetBstMidEndSepPunct{\mcitedefaultmidpunct}
{\mcitedefaultendpunct}{\mcitedefaultseppunct}\relax
\EndOfBibitem
\bibitem{Mattson:2002vu}
SELEX collaboration, M.~Mattson {\em et~al.},
  \ifthenelse{\boolean{articletitles}}{\emph{{First observation of the doubly
  charmed baryon \Xiccp}},
  }{}\href{https://doi.org/10.1103/PhysRevLett.89.112001}{Phys.\ Rev.\ Lett.\
  \textbf{89} (2002) 112001},
  \href{http://arxiv.org/abs/hep-ex/0208014}{{\normalfont\ttfamily
  arXiv:hep-ex/0208014}}\relax
\mciteBstWouldAddEndPuncttrue
\mciteSetBstMidEndSepPunct{\mcitedefaultmidpunct}
{\mcitedefaultendpunct}{\mcitedefaultseppunct}\relax
\EndOfBibitem
\bibitem{Ocherashvili:2004hi}
SELEX collaboration, A.~Ocherashvili {\em et~al.},
  \ifthenelse{\boolean{articletitles}}{\emph{{Confirmation of the double charm
  baryon $\Xiccp(3520)$ via its decay to $p\Dp\Km$}},
  }{}\href{https://doi.org/10.1016/j.physletb.2005.09.043}{Phys.\ Lett.\
  \textbf{B628} (2005) 18},
  \href{http://arxiv.org/abs/hep-ex/0406033}{{\normalfont\ttfamily
  arXiv:hep-ex/0406033}}\relax
\mciteBstWouldAddEndPuncttrue
\mciteSetBstMidEndSepPunct{\mcitedefaultmidpunct}
{\mcitedefaultendpunct}{\mcitedefaultseppunct}\relax
\EndOfBibitem
\bibitem{Karliner:2017gml}
M.~Karliner and J.~L. Rosner,
  \ifthenelse{\boolean{articletitles}}{\emph{{Isospin splittings in baryons
  with two heavy quarks}},
  }{}\href{https://doi.org/10.1103/PhysRevD.96.033004}{Phys.\ Rev.\
  \textbf{D96} (2017) 033004},
  \href{http://arxiv.org/abs/1706.06961}{{\normalfont\ttfamily
  arXiv:1706.06961}}\relax
\mciteBstWouldAddEndPuncttrue
\mciteSetBstMidEndSepPunct{\mcitedefaultmidpunct}
{\mcitedefaultendpunct}{\mcitedefaultseppunct}\relax
\EndOfBibitem
\bibitem{ratti2003new}
S.~P. Ratti, \ifthenelse{\boolean{articletitles}}{\emph{{New results on
  c-baryons and a search for cc-baryons in FOCUS}},
  }{}\href{https://doi.org/10.1016/S0920-5632(02)01948-5}{Nucl.\ Phys.\ Proc.\
  Suppl.\  \textbf{115} (2003) 33}\relax
\mciteBstWouldAddEndPuncttrue
\mciteSetBstMidEndSepPunct{\mcitedefaultmidpunct}
{\mcitedefaultendpunct}{\mcitedefaultseppunct}\relax
\EndOfBibitem
\bibitem{Aubert:2006qw}
BaBar collaboration, B.~Aubert {\em et~al.},
  \ifthenelse{\boolean{articletitles}}{\emph{{Search for doubly charmed baryons
  \Xiccp and \Xiccpp in BaBar}},
  }{}\href{https://doi.org/10.1103/PhysRevD.74.011103}{Phys.\ Rev.\
  \textbf{D74} (2006) 011103},
  \href{http://arxiv.org/abs/hep-ex/0605075}{{\normalfont\ttfamily
  arXiv:hep-ex/0605075}}\relax
\mciteBstWouldAddEndPuncttrue
\mciteSetBstMidEndSepPunct{\mcitedefaultmidpunct}
{\mcitedefaultendpunct}{\mcitedefaultseppunct}\relax
\EndOfBibitem
\bibitem{Chistov:2006zj}
Belle collaboration, R.~Chistov {\em et~al.},
  \ifthenelse{\boolean{articletitles}}{\emph{{Observation of new states
  decaying into \Lcp\Km\pip and \Lcp\KS\pim}},
  }{}\href{https://doi.org/10.1103/PhysRevLett.97.162001}{Phys.\ Rev.\ Lett.\
  \textbf{97} (2006) 162001},
  \href{http://arxiv.org/abs/hep-ex/0606051}{{\normalfont\ttfamily
  arXiv:hep-ex/0606051}}\relax
\mciteBstWouldAddEndPuncttrue
\mciteSetBstMidEndSepPunct{\mcitedefaultmidpunct}
{\mcitedefaultendpunct}{\mcitedefaultseppunct}\relax
\EndOfBibitem
\bibitem{LHCb-PAPER-2013-049}
LHCb collaboration, R.~Aaij {\em et~al.},
  \ifthenelse{\boolean{articletitles}}{\emph{{Search for the doubly charmed
  baryon $\Xires_{cc}^+$}},
  }{}\href{https://doi.org/10.1007/JHEP12(2013)090}{JHEP \textbf{12} (2013)
  090}, \href{http://arxiv.org/abs/1310.2538}{{\normalfont\ttfamily
  arXiv:1310.2538}}\relax
\mciteBstWouldAddEndPuncttrue
\mciteSetBstMidEndSepPunct{\mcitedefaultmidpunct}
{\mcitedefaultendpunct}{\mcitedefaultseppunct}\relax
\EndOfBibitem
\bibitem{LHCb-DP-2019-001}
R.~Aaij {\em et~al.}, \ifthenelse{\boolean{articletitles}}{\emph{{Performance
  of the LHCb trigger and full real-time reconstruction in Run 2 of the LHC}},
  }{}\href{https://doi.org/10.1088/1748-0221/14/04/P04013}{JINST \textbf{14}
  (2019) P04013}, \href{http://arxiv.org/abs/1812.10790}{{\normalfont\ttfamily
  arXiv:1812.10790}}\relax
\mciteBstWouldAddEndPuncttrue
\mciteSetBstMidEndSepPunct{\mcitedefaultmidpunct}
{\mcitedefaultendpunct}{\mcitedefaultseppunct}\relax
\EndOfBibitem
\bibitem{Yu:2017zst}
F.-S. Yu {\em et~al.}, \ifthenelse{\boolean{articletitles}}{\emph{{Discovery
  Potentials of Doubly Charmed Baryons}},
  }{}\href{https://doi.org/10.1088/1674-1137/42/5/051001}{Chin.\ Phys.\
  \textbf{C42} (2018) 051001},
  \href{http://arxiv.org/abs/1703.09086}{{\normalfont\ttfamily
  arXiv:1703.09086}}\relax
\mciteBstWouldAddEndPuncttrue
\mciteSetBstMidEndSepPunct{\mcitedefaultmidpunct}
{\mcitedefaultendpunct}{\mcitedefaultseppunct}\relax
\EndOfBibitem
\bibitem{pdg18}
Particle Data Group, M.~Tanabashi {\em et~al.},
  \ifthenelse{\boolean{articletitles}}{\emph{{Review of Particle Physics}},
  }{}\href{https://doi.org/10.1103/PhysRevD.98.030001}{Phys.\ Rev.\
  \textbf{D98} (2018) 030001}\relax
\mciteBstWouldAddEndPuncttrue
\mciteSetBstMidEndSepPunct{\mcitedefaultmidpunct}
{\mcitedefaultendpunct}{\mcitedefaultseppunct}\relax
\EndOfBibitem
\bibitem{Alves:2008zz}
LHCb collaboration, A.~A. Alves~Jr.\ {\em et~al.},
  \ifthenelse{\boolean{articletitles}}{\emph{{The \lhcb detector at the LHC}},
  }{}\href{https://doi.org/10.1088/1748-0221/3/08/S08005}{JINST \textbf{3}
  (2008) S08005}\relax
\mciteBstWouldAddEndPuncttrue
\mciteSetBstMidEndSepPunct{\mcitedefaultmidpunct}
{\mcitedefaultendpunct}{\mcitedefaultseppunct}\relax
\EndOfBibitem
\bibitem{LHCb-DP-2014-002}
LHCb collaboration, R.~Aaij {\em et~al.},
  \ifthenelse{\boolean{articletitles}}{\emph{{LHCb detector performance}},
  }{}\href{https://doi.org/10.1142/S0217751X15300227}{Int.\ J.\ Mod.\ Phys.\
  \textbf{A30} (2015) 1530022},
  \href{http://arxiv.org/abs/1412.6352}{{\normalfont\ttfamily
  arXiv:1412.6352}}\relax
\mciteBstWouldAddEndPuncttrue
\mciteSetBstMidEndSepPunct{\mcitedefaultmidpunct}
{\mcitedefaultendpunct}{\mcitedefaultseppunct}\relax
\EndOfBibitem
\bibitem{LHCb-DP-2014-001}
R.~Aaij {\em et~al.}, \ifthenelse{\boolean{articletitles}}{\emph{{Performance
  of the LHCb Vertex Locator}},
  }{}\href{https://doi.org/10.1088/1748-0221/9/09/P09007}{JINST \textbf{9}
  (2014) P09007}, \href{http://arxiv.org/abs/1405.7808}{{\normalfont\ttfamily
  arXiv:1405.7808}}\relax
\mciteBstWouldAddEndPuncttrue
\mciteSetBstMidEndSepPunct{\mcitedefaultmidpunct}
{\mcitedefaultendpunct}{\mcitedefaultseppunct}\relax
\EndOfBibitem
\bibitem{LHCb-DP-2017-001}
P.~d'Argent {\em et~al.}, \ifthenelse{\boolean{articletitles}}{\emph{{Improved
  performance of the LHCb Outer Tracker in LHC Run 2}},
  }{}\href{https://doi.org/10.1088/1748-0221/12/11/P11016}{JINST \textbf{12}
  (2017) P11016}, \href{http://arxiv.org/abs/1708.00819}{{\normalfont\ttfamily
  arXiv:1708.00819}}\relax
\mciteBstWouldAddEndPuncttrue
\mciteSetBstMidEndSepPunct{\mcitedefaultmidpunct}
{\mcitedefaultendpunct}{\mcitedefaultseppunct}\relax
\EndOfBibitem
\bibitem{LHCb-DP-2012-003}
M.~Adinolfi {\em et~al.},
  \ifthenelse{\boolean{articletitles}}{\emph{{Performance of the \lhcb RICH
  detector at the LHC}},
  }{}\href{https://doi.org/10.1140/epjc/s10052-013-2431-9}{Eur.\ Phys.\ J.\
  \textbf{C73} (2013) 2431},
  \href{http://arxiv.org/abs/1211.6759}{{\normalfont\ttfamily
  arXiv:1211.6759}}\relax
\mciteBstWouldAddEndPuncttrue
\mciteSetBstMidEndSepPunct{\mcitedefaultmidpunct}
{\mcitedefaultendpunct}{\mcitedefaultseppunct}\relax
\EndOfBibitem
\bibitem{LHCb-DP-2012-002}
A.~A. Alves~Jr.\ {\em et~al.},
  \ifthenelse{\boolean{articletitles}}{\emph{{Performance of the LHCb muon
  system}}, }{}\href{https://doi.org/10.1088/1748-0221/8/02/P02022}{JINST
  \textbf{8} (2013) P02022},
  \href{http://arxiv.org/abs/1211.1346}{{\normalfont\ttfamily
  arXiv:1211.1346}}\relax
\mciteBstWouldAddEndPuncttrue
\mciteSetBstMidEndSepPunct{\mcitedefaultmidpunct}
{\mcitedefaultendpunct}{\mcitedefaultseppunct}\relax
\EndOfBibitem
\bibitem{LHCb-DP-2016-001}
R.~Aaij {\em et~al.}, \ifthenelse{\boolean{articletitles}}{\emph{{Tesla: an
  application for real-time data analysis in High Energy Physics}},
  }{}\href{https://doi.org/10.1016/j.cpc.2016.07.022}{Comput.\ Phys.\ Commun.\
  \textbf{208} (2016) 35},
  \href{http://arxiv.org/abs/1604.05596}{{\normalfont\ttfamily
  arXiv:1604.05596}}\relax
\mciteBstWouldAddEndPuncttrue
\mciteSetBstMidEndSepPunct{\mcitedefaultmidpunct}
{\mcitedefaultendpunct}{\mcitedefaultseppunct}\relax
\EndOfBibitem
\bibitem{Sjostrand:2007gs}
T.~Sj\"{o}strand, S.~Mrenna, and P.~Skands,
  \ifthenelse{\boolean{articletitles}}{\emph{{A brief introduction to PYTHIA
  8.1}}, }{}\href{https://doi.org/10.1016/j.cpc.2008.01.036}{Comput.\ Phys.\
  Commun.\  \textbf{178} (2008) 852},
  \href{http://arxiv.org/abs/0710.3820}{{\normalfont\ttfamily
  arXiv:0710.3820}}\relax
\mciteBstWouldAddEndPuncttrue
\mciteSetBstMidEndSepPunct{\mcitedefaultmidpunct}
{\mcitedefaultendpunct}{\mcitedefaultseppunct}\relax
\EndOfBibitem
\bibitem{Sjostrand:2006za}
T.~Sj\"{o}strand, S.~Mrenna, and P.~Skands,
  \ifthenelse{\boolean{articletitles}}{\emph{{PYTHIA 6.4 physics and manual}},
  }{}\href{https://doi.org/10.1088/1126-6708/2006/05/026}{JHEP \textbf{05}
  (2006) 026}, \href{http://arxiv.org/abs/hep-ph/0603175}{{\normalfont\ttfamily
  arXiv:hep-ph/0603175}}\relax
\mciteBstWouldAddEndPuncttrue
\mciteSetBstMidEndSepPunct{\mcitedefaultmidpunct}
{\mcitedefaultendpunct}{\mcitedefaultseppunct}\relax
\EndOfBibitem
\bibitem{LHCb-PROC-2010-056}
I.~Belyaev {\em et~al.}, \ifthenelse{\boolean{articletitles}}{\emph{{Handling
  of the generation of primary events in Gauss, the LHCb simulation
  framework}}, }{}\href{https://doi.org/10.1088/1742-6596/331/3/032047}{J.\
  Phys.\ Conf.\ Ser.\  \textbf{331} (2011) 032047}\relax
\mciteBstWouldAddEndPuncttrue
\mciteSetBstMidEndSepPunct{\mcitedefaultmidpunct}
{\mcitedefaultendpunct}{\mcitedefaultseppunct}\relax
\EndOfBibitem
\bibitem{Chang:2009va}
C.-H. Chang, J.-X. Wang, and X.-G. Wu,
  \ifthenelse{\boolean{articletitles}}{\emph{{GENXICC2.0: An upgraded version
  of the generator for hadronic production of double heavy baryons
  $\Xires_{cc}$, $\Xires_{bc}$ and $\Xires_{bb}$}},
  }{}\href{https://doi.org/10.1016/j.cpc.2010.02.008}{Comput.\ Phys.\ Commun.\
  \textbf{181} (2010) 1144},
  \href{http://arxiv.org/abs/0910.4462}{{\normalfont\ttfamily
  arXiv:0910.4462}}\relax
\mciteBstWouldAddEndPuncttrue
\mciteSetBstMidEndSepPunct{\mcitedefaultmidpunct}
{\mcitedefaultendpunct}{\mcitedefaultseppunct}\relax
\EndOfBibitem
\bibitem{Lange:2001uf}
D.~J. Lange, \ifthenelse{\boolean{articletitles}}{\emph{{The EvtGen particle
  decay simulation package}},
  }{}\href{https://doi.org/10.1016/S0168-9002(01)00089-4}{Nucl.\ Instrum.\
  Meth.\  \textbf{A462} (2001) 152}\relax
\mciteBstWouldAddEndPuncttrue
\mciteSetBstMidEndSepPunct{\mcitedefaultmidpunct}
{\mcitedefaultendpunct}{\mcitedefaultseppunct}\relax
\EndOfBibitem
\bibitem{Golonka:2005pn}
P.~Golonka and Z.~Was, \ifthenelse{\boolean{articletitles}}{\emph{{PHOTOS Monte
  Carlo: A precision tool for QED corrections in $Z$ and $W$ decays}},
  }{}\href{https://doi.org/10.1140/epjc/s2005-02396-4}{Eur.\ Phys.\ J.\
  \textbf{C45} (2006) 97},
  \href{http://arxiv.org/abs/hep-ph/0506026}{{\normalfont\ttfamily
  arXiv:hep-ph/0506026}}\relax
\mciteBstWouldAddEndPuncttrue
\mciteSetBstMidEndSepPunct{\mcitedefaultmidpunct}
{\mcitedefaultendpunct}{\mcitedefaultseppunct}\relax
\EndOfBibitem
\bibitem{Allison:2006ve}
Geant4 collaboration, J.~Allison {\em et~al.},
  \ifthenelse{\boolean{articletitles}}{\emph{{Geant4 developments and
  applications}}, }{}\href{https://doi.org/10.1109/TNS.2006.869826}{IEEE
  Trans.\ Nucl.\ Sci.\  \textbf{53} (2006) 270}\relax
\mciteBstWouldAddEndPuncttrue
\mciteSetBstMidEndSepPunct{\mcitedefaultmidpunct}
{\mcitedefaultendpunct}{\mcitedefaultseppunct}\relax
\EndOfBibitem
\bibitem{Agostinelli:2002hh}
Geant4 collaboration, S.~Agostinelli {\em et~al.},
  \ifthenelse{\boolean{articletitles}}{\emph{{Geant4: A simulation toolkit}},
  }{}\href{https://doi.org/10.1016/S0168-9002(03)01368-8}{Nucl.\ Instrum.\
  Meth.\  \textbf{A506} (2003) 250}\relax
\mciteBstWouldAddEndPuncttrue
\mciteSetBstMidEndSepPunct{\mcitedefaultmidpunct}
{\mcitedefaultendpunct}{\mcitedefaultseppunct}\relax
\EndOfBibitem
\bibitem{LHCb-PROC-2011-006}
M.~Clemencic {\em et~al.}, \ifthenelse{\boolean{articletitles}}{\emph{{The
  \lhcb simulation application, Gauss: Design, evolution and experience}},
  }{}\href{https://doi.org/10.1088/1742-6596/331/3/032023}{J.\ Phys.\ Conf.\
  Ser.\  \textbf{331} (2011) 032023}\relax
\mciteBstWouldAddEndPuncttrue
\mciteSetBstMidEndSepPunct{\mcitedefaultmidpunct}
{\mcitedefaultendpunct}{\mcitedefaultseppunct}\relax
\EndOfBibitem
\bibitem{Hocker:2007ht}
H.~Voss, A.~Hoecker, J.~Stelzer, and F.~Tegenfeldt,
  \ifthenelse{\boolean{articletitles}}{\emph{{TMVA - Toolkit for Multivariate
  Data Analysis with ROOT}}, }{}\href{https://doi.org/10.22323/1.050.0040}{PoS
  \textbf{ACAT} (2007) 040}\relax
\mciteBstWouldAddEndPuncttrue
\mciteSetBstMidEndSepPunct{\mcitedefaultmidpunct}
{\mcitedefaultendpunct}{\mcitedefaultseppunct}\relax
\EndOfBibitem
\bibitem{Hulsbergen:2005pu}
W.~D. Hulsbergen, \ifthenelse{\boolean{articletitles}}{\emph{{Decay chain
  fitting with a Kalman filter}},
  }{}\href{https://doi.org/10.1016/j.nima.2005.06.078}{Nucl.\ Instrum.\ Meth.\
  \textbf{A552} (2005) 566},
  \href{http://arxiv.org/abs/physics/0503191}{{\normalfont\ttfamily
  arXiv:physics/0503191}}\relax
\mciteBstWouldAddEndPuncttrue
\mciteSetBstMidEndSepPunct{\mcitedefaultmidpunct}
{\mcitedefaultendpunct}{\mcitedefaultseppunct}\relax
\EndOfBibitem
\bibitem{Punzi:2003bu}
G.~Punzi, \ifthenelse{\boolean{articletitles}}{\emph{{Sensitivity of searches
  for new signals and its optimization}}, }{}eConf \textbf{C030908} (2003)
  MODT002, \href{http://arxiv.org/abs/physics/0308063}{{\normalfont\ttfamily
  arXiv:physics/0308063}}\relax
\mciteBstWouldAddEndPuncttrue
\mciteSetBstMidEndSepPunct{\mcitedefaultmidpunct}
{\mcitedefaultendpunct}{\mcitedefaultseppunct}\relax
\EndOfBibitem
\bibitem{Skwarnicki:1986xj}
T.~Skwarnicki, {\em {A study of the radiative cascade transitions between the
  Upsilon-prime and Upsilon resonances}}, PhD thesis, Institute of Nuclear
  Physics, Krakow, 1986,
  {\href{http://inspirehep.net/record/230779/}{DESY-F31-86-02}}\relax
\mciteBstWouldAddEndPuncttrue
\mciteSetBstMidEndSepPunct{\mcitedefaultmidpunct}
{\mcitedefaultendpunct}{\mcitedefaultseppunct}\relax
\EndOfBibitem
\bibitem{Cowan:2010js}
G.~Cowan, K.~Cranmer, E.~Gross, and O.~Vitells,
  \ifthenelse{\boolean{articletitles}}{\emph{{Asymptotic formulae for
  likelihood-based tests of new physics}},
  }{}\href{https://doi.org/10.1140/epjc/s10052-011-1554-0}{Eur.\ Phys.\ J.\
  \textbf{C71} (2011) 1554}, Erratum
  \href{https://doi.org/10.1140/epjc/s10052-013-2501-z}{ibid.\   \textbf{C73}
  (2013) 2501}, \href{http://arxiv.org/abs/1007.1727}{{\normalfont\ttfamily
  arXiv:1007.1727}}\relax
\mciteBstWouldAddEndPuncttrue
\mciteSetBstMidEndSepPunct{\mcitedefaultmidpunct}
{\mcitedefaultendpunct}{\mcitedefaultseppunct}\relax
\EndOfBibitem
\bibitem{LHCb-PUB-2016-021}
L.~Anderlini {\em et~al.}, \ifthenelse{\boolean{articletitles}}{\emph{{The
  PIDCalib package}}, }{}
  \href{http://cdsweb.cern.ch/search?p=LHCb-PUB-2016-021&f=reportnumber&action_search=Search&c=LHCb+Notes}
  {LHCb-PUB-2016-021}, 2016\relax
\mciteBstWouldAddEndPuncttrue
\mciteSetBstMidEndSepPunct{\mcitedefaultmidpunct}
{\mcitedefaultendpunct}{\mcitedefaultseppunct}\relax
\EndOfBibitem
\bibitem{Read:2002hq}
A.~L. Read, \ifthenelse{\boolean{articletitles}}{\emph{{Presentation of search
  results: The CL(s) technique}},
  }{}\href{https://doi.org/10.1088/0954-3899/28/10/313}{J.\ Phys.\
  \textbf{G28} (2002) 2693}\relax
\mciteBstWouldAddEndPuncttrue
\mciteSetBstMidEndSepPunct{\mcitedefaultmidpunct}
{\mcitedefaultendpunct}{\mcitedefaultseppunct}\relax
\EndOfBibitem
\end{mcitethebibliography}
 
\newpage
\centerline
{\large\bf LHCb Collaboration}
\begin
{flushleft}
\small
R.~Aaij$^{29}$,
C.~Abell{\'a}n~Beteta$^{46}$,
B.~Adeva$^{43}$,
M.~Adinolfi$^{50}$,
C.A.~Aidala$^{77}$,
Z.~Ajaltouni$^{7}$,
S.~Akar$^{61}$,
P.~Albicocco$^{20}$,
J.~Albrecht$^{12}$,
F.~Alessio$^{44}$
% ...

{\footnotesize \it

% ...
$ ^{7}$Universit{\'e} Clermont Auvergne, CNRS/IN2P3, LPC, Clermont-Ferrand, France\\
% ...
$ ^{12}$Fakult{\"a}t Physik, Technische Universit{\"a}t Dortmund, Dortmund, Germany\\
% ...
$ ^{20}$INFN Laboratori Nazionali di Frascati, Frascati, Italy\\
% ...
$ ^{29}$Nikhef National Institute for Subatomic Physics, Amsterdam, Netherlands\\
% ...
$ ^{43}$Instituto Galego de F{\'\i}sica de Altas Enerx{\'\i}as (IGFAE), Universidade de Santiago de Compostela, Santiago de Compostela, Spain\\
$ ^{44}$European Organization for Nuclear Research (CERN), Geneva, Switzerland\\
% ...
$ ^{46}$Physik-Institut, Universit{\"a}t Z{\"u}rich, Z{\"u}rich, Switzerland\\
% ...
$ ^{50}$H.H. Wills Physics Laboratory, University of Bristol, Bristol, United Kingdom\\
% ...
$ ^{61}$University of Cincinnati, Cincinnati, OH, United States\\
% ...
$ ^{63}$Syracuse University, Syracuse, NY, United States\\
% ...
$ ^{77}$University of Michigan, Ann Arbor, United States, associated to $^{63}$\\
% ...
}
\end{flushleft}

\end{document}